\documentclass{JHEP3}

\usepackage{epsfig,multicol,bbm,amsmath,amssymb,euscript,array,,mathrsfs,amsfonts}

\title{Unified Dark Matter Scalar Field Models }
\author{Daniele Bertacca$^{a,b,c}$,  Nicola Bartolo$^{a,b}$, Sabino Matarrese$^{a,b}$\\
$^a$ Dipartimento di Fisica Galileo Galilei  Universit\`{a} di Padova, via F. Marzolo, 8 I-35131 Padova, Italy\\
$^b$ INFN Sezione di Padova, via F. Marzolo, 8 I-35131 Padova, Italy\\
$^c$ Institute of Cosmology \& Gravitation, University of Portsmouth,
Dennis Sciama Building, Portsmouth, PO1 3FX, United Kingdom\\
E-mails: \email{daniele.bertacca@pd.infn.it, daniele.bertacca@port.ac.uk}, \email{nicola.bartolo@pd.infn.it}, \email{sabino.matarrese@pd.infn.it}}

\abstract{In this work we analyze and review cosmological models in which the 
dynamics of a single scalar field accounts for a unified description 
of the Dark Matter and Dark Energy sectors, dubbed Unified Dark Matter (UDM) models.
In this framework, we consider the general Lagrangian of 
k-\emph{essence}, which allows to find solutions around 
which the scalar field describes the desired mixture of Dark Matter and Dark Energy.\\  
We also discuss static and spherically symmetric 
solutions of Einstein's equations for a scalar field 
with non-canonical kinetic term, in connection with galactic halo rotation curves.}

\begin{document}

\section{Introduction}

\normalsize
In the last few decades a standard cosmological ``Big Bang" model has emerged,
based on Einstein's theory of gravity, General Relativity.
Indeed, observations tell us that - by and large -  the Universe looks the same
 in all directions, and it is assumed to be homogeneous on the basis of the
``Cosmological Principle", i.e.\ a cosmological version of the  Copernican principle.
The request for the Universe to be homogeneous and isotropic translates, in the language
of space-time, in a Robertson-Walker metric. Assuming the latter, Einstein equations
simplify, becoming the Friedmann equations, and in general the solutions of these equations
are called Friedmann-Lemaitre-Robertson-Walker (FLRW) models.
The cosmological inhomogeneities  we observe on the largest scales as tiny anisotropies of
the  Cosmic Microwave Background (CMB) are then well explained  by small relativistic
perturbations of these FLRW ``background" models, while on smaller scales the inhomogeneities
are larger and call for  non-linear dynamics, but relativistic effects are negligible and
Newtonian dynamics is sufficient to explain the formation of the structures we see,
i.e.\ galaxies, groups and clusters forming the observed ``cosmic web".
In this context, last decade's  observations of large scale structure, search for
Ia supernovae (SNIa) \cite{Perlmutter:1998np, Riess:1998cb, Riess:1998dv, Amanullah:2010vv} and measurements of the CMB anisotropies
\cite{Larson:2010gs, Komatsu:2010fb} suggest that two dark
components govern the dynamics of the Universe. They are the dark matter (DM), thought to be
the main responsible for structure formation, and an additional dark energy (DE) component
that is supposed to drive the measured cosmic acceleration \cite{Tsujikawa:2010sc,Copeland:2006wr}.
However, the DM particles have
not yet been detected in the lab, although there are hints for their existence from cosmic
rays experiments \cite{Adriani:2008zr, Adriani:2008zq, Bernabei:2000qi},
and there is no theoretical justification for the tiny
cosmological constant \cite{Weinberg:1988cp} (or more general DE component\cite{Tsujikawa:2010sc,Copeland:2006wr})
implied by observations (see also \cite{Amendola:1272934}).
Therefore, over the last decade, the search for extended theories of
gravity has flourished as a possible alternative to DE \cite{Tsujikawa:2010sc,Copeland:2006wr}.
At the same time, in the context of General Relativity,
it is very interesting to study the possibility of  an interaction between Dark Matter
and Dark Energy without violating current observational constraints \cite{Tsujikawa:2010sc,Copeland:2006wr,Baldi:2010vv,Baldi:2008ay,
Valiviita:2008iv,CalderaCabral:2008bx,CalderaCabral:2009ja,Quercellini:2008vh} (see also \cite{Leon:2009dt}).
This possibility  could alleviate the so called ``coincidence problem", namely, why are the
energy densities of the two dark components of the same order of magnitude today.
Another more radical  explanation of  the observed cosmic
acceleration and structure formation  is to assume the existence
of a single dark component: Unified Dark Matter (UDM) models, see e.g.\ \cite{Sahni:1999qe, Kamenshchik:2001cp,Bilic:2001cg,Bento:2002ps,Carturan:2002si, Padmanabhan:2002sh, Sandvik:2002jz, Amendola:2003bz, Makler:2003iw, Scherrer:2004au, Giannakis:2005kr, Bertacca:2007cv, Bertacca:2007ux, Bertacca:2007fc,Chimento:2007da, Quercellini:2007ht, Balbi:2007mz, Bertacca:2008uf, Pietrobon:2008js, Bilic:2008yr, Camera:2009uz, Li:2009mf, Chimento:2009nj, Piattella:2009kt, Gao:2009me, Camera:2010wm, Lim:2010yk, Bertacca-2010-1, Bertacca:2010mt, transfer-function} (see also \cite{Liddle:2006qz, Cardenas:2007xh, Panotopoulos:2007ri, Liddle:2008bm, Bose:2008ew} on how to unify DM, DE, and inflation, \cite{Takahashi:2005kp} on unification of DM and DE in the framework of supersymmetry, \cite{Mainini:2004he, Mainini:2005mq, Mainini:2007nq, LaVacca:2009yp} on unification of DM and DE from the solution of the strong CP-problem, \cite{LukesGerakopoulos:2008rr, Basilakos:2008jb} on unification of DM and DE  in connection with chaotic scalar field solutions in Friedmann-Robertson-Walker cosmologies, \cite{Arbey:2001jj, Arbey:2003sj, Arbey:2006it} on how to unify dark energy and dark matter through a complex scalar field, and \cite{Wetterich:1994bg,Wetterich:2001cn,Wetterich:2001yw} on a study of a scalar field, ``Cosmos Dark Matter'', that induces a time-dependent cosmological ÒconstantÓ).

In comparison with the standard DM + DE models (e.g.\ even the
simplest model, with DM  and a cosmological constant), these models have 
the advantage that we can describe the dynamics of the Universe with a single scalar field which
triggers both the accelerated expansion at late times and the LSS formation at earlier times. 
Specifically, for these models, we can use Lagrangians with a non-canonical kinetic term, 
namely a term which is an arbitrary function of the square of the time derivative of the scalar field,
in the homogeneous and isotropic background.

Originally this method was proposed to have inflation driven by kinetic energy, called $k$-inflation 
\cite{ArmendarizPicon:1999rj,Garriga:1999vw}, to explain early Universe's inflation at high energies.
Then this scenario was applied to DE \cite{Chiba:1999ka, dePutter:2007ny,Linder:2008ya}.
In particular, the analysis was extended to a more general Lagrangian \cite{ArmendarizPicon:2000dh,ArmendarizPicon:2000ah} 
and this scenario was called $k$-essence (see also \cite{Chiba:1999ka,Vikman:2004dc, Rendall:2005fv,Babichev:2007dw, Babichev:2006cy,
Babichev:2007tn,Calcagni:2006ge,Linder:2008ya,Akhoury:2008nn,Ahn:2009xd,Ahn:2009hu, Arroja:2010wy, Unnikrishnan:2010ag}).

For UDM models, several adiabatic or, equivalently, purely kinetic models have been investigated 
in the literature. For example, the generalised Chaplygin gas \cite{Kamenshchik:2001cp, Bilic:2001cg, Bento:2002ps} (see also \cite{Makler:2002jv,Carturan:2002si, Sandvik:2002jz, Bento:2002yx, Alcaniz:2002yt, Avelino:2003cf, Makler:2003iw, Amendola:2003bz, Sen:2005sk, Santos:2006ce, Giannantonio:2006ij, Gorini:2007ta}), the Scherrer \cite{Scherrer:2004au} and generalised Scherrer solutions \cite{Bertacca:2007ux}, the single dark perfect fluid with ``affine'' 2-parameter barotropic equation of state (see \cite{Balbi:2007mz,Pietrobon:2008js} and the corresponding scalar field models \cite{Quercellini:2007ht}) and the homogeneous scalar field deduced from the galactic halo space-time \cite{DiezTejedor:2006qh, Bertacca:2007fc}.
In general, in order for UDM models to  have a background evolution that fits observations and a very small speed of sound, a severe fine-tuning of 
their parameters is necessary (see for example \cite{Pietrobon:2008js, Makler:2003iw, Carturan:2002si, Amendola:2003bz, Sandvik:2002jz, 
Scherrer:2004au, Giannakis:2005kr, Piattella:2009da}).
Finally, one could also easily reinterpret UDM models based on a scalar field Lagrangian in terms of generally 
non-adiabatic fluids \cite{DiezTejedor:2005fz, Brown:1992kc} (see also \cite{Bertacca:2007ux, Bertacca:2008uf}). 
For these models the effective speed of sound, which remains defined in the context of linear perturbation theory, 
is not the same as the adiabatic speed of sound (see \cite{Hu:1998kj}, \cite{Garriga:1999vw} and \cite{Mukhanov:2005sc}).
In \cite{Bertacca:2008uf} a reconstruction technique is devised for the Lagrangian, which allows  to find models 
where the effective speed of sound is small enough, such that the {\it k}-essence scalar field can cluster (see also 
\cite{Camera:2009uz,Camera:2010wm, Bertacca-2010-1, Bertacca:2010mt, transfer-function}).

One of the main issues of these UDM models is whether the single dark fluid is able
to cluster and produce the cosmic structures we observe in the Universe today.
 In fact, a general feature of UDM models is the appearance of an effective sound speed, 
 which may become significantly different from zero during the evolution of the Universe. In general, 
 this corresponds to the appearance of a Jeans length (or sound horizon) below which the dark fluid does not cluster. 
 Thus, the viability of UDM models strictly depends on the value of this effective sound speed 
 \cite{Hu:1998kj,Garriga:1999vw,Mukhanov:2005sc}, which has to be small enough to allow 
 structure formation \cite{Sandvik:2002jz,Giannakis:2005kr,Bertacca:2007cv} and to reproduce 
 the observed pattern of the CMB temperature anisotropies \cite{Carturan:2002si,Bertacca:2007cv}.

In general, in order for UDM models to have a very small speed of sound and a background evolution 
that fits the observations, a severe fine tuning of their parameters is necessary. 
In order to avoid this fine tuning, alternative models with similar goals have been analyzed in the 
literature. Ref.~\cite{Piattella:2009kt} studied in detail the functional form of the Jeans scale in adiabatic 
UDM perturbations and introduced a class of models with a fast transition between an early 
Einstein-de Sitter cold DM-like era and a later $\Lambda$CDM-like phase. 
If the transition is fast enough, these models may exhibit satisfactory structure formation and CMB fluctuations, 
thus presenting a small Jeans length even in the case of a non-negligible sound speed. Ref.~\cite{Gao:2009me} 
explored unification of DM and DE in a theory containing a scalar field of non-Lagrangian type, 
obtained by direct insertion of a kinetic term into the energy-momentum tensor. Finally, Ref.~\cite{Lim:2010yk}
introduced a class of field theories where comprises two scalar fields, one of which is a Lagrange multiplier 
enforcing a constraint between the otherÕs field value and derivative in order to have the sound speed is always identically zero on all backgrounds.

This work is organized as follows.
In Section \ref{Int-UDM}, considering the general Lagrangian of 
k-\emph{essence} models, we layout the basic equations.
In Section \ref{ISW} we present an analytical study of the Integrated 
Sachs-Wolfe (ISW) effect within the framework of UDM. Computing the temperature power spectrum 
of the Cosmic Microwave Background anisotropies one is able to isolate 
those contributions that can potentially lead to strong deviations from 
the usual ISW effect occurring in a $\Lambda$CDM Universe. 
This helps to highlight the crucial role played by the sound speed in the 
unified dark matter models. Our treatment is completely general in that 
all the results depend only on the speed of sound of the dark 
component and thus it can be applied to a variety of unified models, including 
those which are not described by a scalar field but relies on a single 
dark fluid; see also \cite{Bertacca:2007cv}.
In Section \ref{pkUDM} we study and classify UDM models defined by the purely 
kinetic model. We show that these models have only one late-time attractor with 
equation of state equal to minus one (cosmological constant). 
Studying all possible solutions near the attractor which 
describes a unified dark matter fluid; see also  \cite{Bertacca:2007ux}. 
Subsequently, noting that purely kinetic models can be described as adiabatic single fluid, 
for these Lagrangians it is natural to give a graphical description on pressure - energy density plane, (see also \cite{Piattella:2009kt}).
In Section \ref{standardSF},  we present the simplest case of a scalar field with canonical kinetic term 
which unavoidably leads to an effective sound speed equal to the speed of light.
In Section \ref{non-standardSF}, making the stronger assumption 
that the scalar field Lagrangian is exactly constant along solutions of the equation of motion,
we find a general class of k-\emph{essence} 
models whose classical trajectories directly describe a unified Dark 
Matter$/$Dark Energy (cosmological constant) fluid. In particular we consider more general models allow for the 
possibility that the speed of sound is small during  Einstein--de\! Sitter CDM-like era. 
In Section \ref{recUDM}, we investigate the class of UDM models studied In 
Ref.~\cite{Bertacca:2008uf}, which designed a reconstruction technique of the Lagrangian, 
allowing one to find models where the effective speed of sound is small enough, and the 
$k$-essence scalar field can cluster (see also \cite{Camera:2009uz, Camera:2010wm, Bertacca-2010-1, transfer-function}). 
In particular, the authors of Ref.~\cite{Bertacca:2008uf} require that the Lagrangian of the scalar field is constant along 
classical trajectories on cosmological scales, in order to obtain a background identical
to the background of the $\Lambda$CDM model.
In Section \ref{GUDM}, we develop and generalize the approach studied in  Ref.~ \cite{Bertacca:2008uf}. 
Specifically, we focus  on scalar-field Lagrangians with non-canonical kinetic term to obtain UDM models 
that can mimic a fluid of dark matter and quintessence-like dark energy, with the aim of  
studying models where the background does not necessarily mimic
the $\Lambda$CDM background, see also \cite{Bertacca:2010mt}.
In Section \ref{HUDM}, we investigate the static and spherically symmetric 
solutions of Einstein's equations for a scalar field with non-canonical 
kinetic term, assumed to provide both the dark matter and dark energy 
components of the Universe; see also \cite{Bertacca:2007fc}. 
We show that there exist suitable scalar field Lagrangians 
that allow to describe the cosmological background evolution and 
the static solutions with a single dark fluid. In Section \ref{conclusion}, we draw our main conclusions. 
Finally, in the Appendix \ref{Static_genScherrer}, for completeness we provide the spherical collapse top-hat 
solution for UDM models based on purely kinetic scalar Þeld Lagrangians, which allow us 
to connect the cosmological solutions to the static conÞgurations.

\section{Unified Dark Matter Scalar field models}\label{Int-UDM}

We start recalling the main equations which are useful for the description of most the UDM models within the framework of k-essence. 

Consider the action
\begin{equation}\label{eq:action}
S = S_{G} + S_{\varphi} =  \int d^4 x \sqrt{-g} \left[\frac{R}{2}+\mathcal{L}(\varphi, X)\right]\;,
\end{equation}
where
\begin{equation}\label{x}
X = -\frac{1}{2}\nabla_\mu \varphi \nabla^\mu \varphi\;.
\end{equation}
where the symbol $\nabla$ denotes covariant differentiation.
We adopt $8\pi G = c^2 = 1$ units and the $(-,+,+,+)$ signature for the metric (Greek indices run over spacetime dimensions, while Latin indices label spatial coordinates).

The stress-energy tensor of the scalar field $\varphi$ has the following form:
\begin{equation}\label{energy-momentum-tensor}
 T^{\varphi}_{\mu \nu } = - \frac{2}{\sqrt{-g}}\frac{\delta S_{\varphi }}{\delta g^{\mu \nu }}=\frac{\partial \mathcal{L}(\varphi ,X)}{\partial X}\nabla_{\mu }\varphi\nabla _{\nu }\varphi +\mathcal{L}(\varphi ,X)g_{\mu \nu }\;,
\end{equation}
and its equation of motion reads 
\begin{equation}
\label{eq-motion}
\nabla^\mu \left[\frac{\partial\mathcal{L}}{\partial(\partial_\mu \varphi)}
\right]=\frac{\partial\mathcal{L}}{\partial \varphi}\;. 
\end{equation}

If $X$ is time-like then $S_{\varphi}$ describes a perfect fluid $T^{\varphi}_{\mu \nu }= (\rho + p)u_{\mu}u_{\nu } + p\,g_{\mu \nu }$, where the pressure is
\begin{equation}
  \label{pressure}
  \mathcal{L} = p(\varphi ,X)\;,
\end{equation}
and the energy density is
\begin{equation}\label{energy-density}
  \rho = \rho(\varphi ,X)= 2X\frac{\partial p(\varphi ,X)}{\partial X}-p(\varphi ,X)\;.
\end{equation}
The four-velocity has the following form.
\begin{equation}
  \label{eq:four-velocity}
  u_{\mu }= \frac{\nabla _{\mu }\varphi }{\sqrt{2X}}\;.
\end{equation}
Assume a flat, homogeneous Friedmann-Lema\^{i}tre-Robertson-Walker (FLRW) background metric, i.e.
\begin{equation}
ds^2=-dt^2+a(t)^2\delta_{ij} dx^i dx^j = a(\eta)^2 (-d\eta^2 + \delta_{ij} dx^i dx^j)\;,
\end{equation}
where $a(t)$ is the scale factor, $\delta_{ij}$ denotes the unit tensor and $\eta$ is the conformal time. 

Assuming that the energy density of the radiation is negligible at the times of interest, and disregarding also the small baryonic component\footnote{Indeed the density of baryons relative is about $4.5 \%$ today and $16.9 \%$ prior to Dark Energy domination in the standard cosmological model \cite{Larson:2010gs, Komatsu:2010fb}.},
 the background evolution of the Universe is completely characterised by the following equations:
\begin{equation}\label{eq_u1}
\mathcal{H}^2 = a^2 H^2 = \frac{1}{3} a^2 \rho\;,
\end{equation}
and
\begin{equation}\label{eq_u2}
\mathcal{H}'-\mathcal{H}^2=a^2 \dot{H} = - \frac{1}{2} a^2(p + \rho)\;,
\end{equation}
where ${\cal H} = a'/a$ and $H = \dot{a}/a$. The dot denotes differentiation with respect to (wrt) the cosmic time $t$ whereas a prime denotes differentiation wrt the conformal time $\eta$.

In the background we have that $X = \dot{\varphi}^2/2=\varphi'^2/(2a^2)$, therefore the equation of motion Eq.~(\ref{eq-motion}) for the homogeneous mode $\varphi(t)$ becomes
\begin{equation}\label{eq_phi}
\left(\frac{\partial p}{\partial X} + 2X\frac{\partial^2 p}{\partial X^2}\right)\ddot\varphi + \frac{\partial p}{\partial X}(3H\dot\varphi) + \frac{\partial^2 p}{\partial \varphi \partial X}\dot\varphi^2 - \frac{\partial p}{\partial \varphi} = 0\;.
\end{equation}
An important quantity is the Equation of State (EoS) parameter $w \equiv p/\rho$, which in our case reads
\begin{equation}\label{w}
w = \frac{p}{2X (\partial p/\partial X) - p}\;.
\end{equation}
We mainly focus on the other relevant physical quantity, the speed of sound, which enters in governing the evolution of the scalar field perturbations. Consider small inhomogeneities of the scalar field, i.e.
\begin{equation}
\varphi(t,\mbox{\boldmath x})= \varphi_0(t)+\delta\varphi(t,\mbox{\boldmath $x$})\;,
\end{equation}
and write the perturbed FLRW metric in the longitudinal gauge as
\begin{equation}
ds^2 = -(1 + 2 \Phi)dt^2 + (1 - 2 \Phi)a(t)^2\delta_{ij} dx^i dx^j\;, 
\end{equation}
being $\delta T_{i}^j = 0$ for $i \neq j$ \cite{Mukhanov:1990me}. The linearised $(0-0)$ and $(0-i)$ Einstein equations are (see Ref.~\cite{Garriga:1999vw} and Ref.~\cite{Mukhanov:2005sc})
\begin{equation}
\delta T^{\varphi \;0}_{\phantom{\varphi \;}0}=-\delta \rho= \frac{\partial  \rho}{\partial \phi} \delta\phi 
- \frac{\partial  \rho}{\partial X}  \delta X=
-\frac{p+\rho}{c_{\rm s}^2}\left[\left(\frac{\delta\varphi}{\varphi_0'} \right)'
+\mathcal{H} \frac{\delta\varphi}{\varphi_0'}- \Phi\right]+
3\mathcal{H}(p+\rho) \frac{\delta\varphi}{\varphi_0'}\;,
\end{equation}
and
\begin{equation}
\delta T^{\varphi \;0}_{\phantom{\varphi \;}i}=-(p+\rho)\left( \frac{\delta\varphi}{\varphi_0'} \right)_{,i}\;,
\end{equation}
where one defines a ``speed of sound'' $c_{\rm s}^2$ relative to the pressure and energy density fluctuation of the kinetic term~\cite{Garriga:1999vw} as follows:
\begin{equation}\label{cs}
c_{\rm s}^2 \equiv  \frac{\partial p /\partial X}{\partial \rho /\partial X} = \frac{\partial p/\partial X}{(\partial p/\partial X)+ 2X(\partial^2p/\partial X^2)} \;. 
\end{equation}
From the above linearized Einstein's equations one obtains \cite{Garriga:1999vw, Mukhanov:2005sc}
\begin{equation}\label{pertu-eq1}
\nabla^2 \Phi = \frac{1}{2} \frac{a^2(p+\rho)}{c_{\rm s}^2 \mathcal{H}} \left(\mathcal{H}  \frac{\delta\varphi}{\varphi_0'}+ \Phi \right)'\;,
\end{equation}
and
\begin{equation}
\label{pertu-eq2}
\left(a^2 \frac{\Phi}{\mathcal{H}} \right)'= \frac{1}{2} \frac{a^2(p+\rho)}
{\mathcal{H}^2}\left(\mathcal{H}  \frac{\delta\varphi}{\varphi_0'}
+ \Phi \right) \;.
\end{equation}
Eqs.~(\ref{pertu-eq1}) and~(\ref{pertu-eq2}) are sufficient to determine the gravitational potential $\Phi$ and the perturbation of the scalar field. It is useful to write explicitly the perturbed scalar field as a function of the gravitational potential
\begin{equation}
\label{delta-phi}
 \frac{\delta\varphi}{\varphi_0'}=2\frac{\Phi'+ \mathcal{H}\Phi}{a^2(p+\rho)}\;.
\end{equation}
Defining two new variables
\begin{equation}
\label{u-v}
u\equiv 2 \frac{\Phi}{(p+\rho)^{1/2}} \;, \quad\quad v\equiv z \left(\mathcal{H}  \frac{\delta\varphi}{\varphi_0'}+ \Phi \right)\;,
\end{equation}
where $z = a^2(p+\rho)^{1/2}/(c_{\rm s} \mathcal{H})$, we can recast (\ref{pertu-eq1}) and (\ref{pertu-eq2}) in terms of $u$ and $v$ \cite{Mukhanov:2005sc}:
\begin{equation}
\label{pertu-eq_uv}
c_{\rm s} \nabla^2 u = z \left( \frac{v}{z}\right)'\;,\quad \quad 
c_{\rm s} v= \theta \left( \frac{u}{\theta}\right)'
\end{equation}
where $\theta = 1/(c_{\rm s} z)=(1+p/\rho)^{-1/2}/(\sqrt{3}a)$. Starting from (\ref{pertu-eq_uv}) we arrive at the following second order differential equations for $u$ \cite{Mukhanov:2005sc}:
\begin{equation}\label{diff-eq_u}
u''- c_{\rm s}^2 \nabla^2 u - \frac{\theta''}{\theta}u = 0\;.
\end{equation}
Unfortunately, we do not know the exact solution for a generic Lagrangian. However, we can consider the asymptotic solutions, i.e. the long-wavelength and the short-wavelength perturbations, depending whether $c_{\rm s}^2k^2 \ll \left|\theta''/\theta\right|$ or $c_{\rm s}^2k^2 \gg \left|\theta''/\theta\right|$, respectively. 

Starting from Eq. (\ref{diff-eq_u}), let us define the squared Jeans wave number \cite{Bertacca:2007cv}:
\begin{equation}
 k^{2}_{\rm J} := \left|\frac{\theta''}{c_{\rm s}^{2}\theta}\right|\;.
\end{equation}
Its reciprocal  defines the squared Jeans length: $\lambda^2_{\rm J} \simeq a^{2}/k^2_{\rm J}$.

There are two regimes of evolution. If $k^2 \gg k_{\rm J}^2$ and the speed of sound is slowly varying, then the solution of Eq. (\ref{diff-eq_u}) is
\begin{equation}\label{ukggkJ}
 u \simeq \frac{C}{\sqrt{c_{\rm s}}}\exp\left(\pm ik\int c_{\rm s}d\eta\right)\;,
\end{equation}
where $C$ is an appropriate integration constant\footnote{This solution is exact if the speed of sound satisfies the equation $2c_{\rm s}''c_{\rm s} - 3\left(c_{\rm s}'\right)^2 = 0$, which implies 
\begin{eqnarray}
 c_{\rm s} = \frac{4}{\left(c_1\eta + c_2\right)^2}\;,\nonumber
\end{eqnarray}
where $c_1$ and $c_2$ are generic constants. A particular case is when $c_1 = 0$, for which the speed of sound is constant.}.
On these scales, smaller than the Jeans length, the gravitational potential oscillates and decays in  time, with observable effects on both the CMB and the matter power spectra \cite{Bertacca:2007cv}.

For large scale perturbations, when $k^2 \ll k_{\rm J}^2$, Eq. (\ref{diff-eq_u}) can be rewritten as  $u''/u  \simeq \theta''/\theta$, with  general solution
\begin{equation}\label{uklesskJ}
 u \simeq \kappa_1\theta + \kappa_2\theta\int \frac{d\eta}{\theta^{2}}\;.
\end{equation}
In this large scale limit the evolution of the gravitational potential $\Phi$ depends only on the background evolution, encoded in $\theta$, i.e.\ it is the same for all $k$ modes. The first term $\kappa_1\theta$ is the usual decaying mode, which we are going to neglect in the following, while $\kappa_2$ is related to the power spectrum, see e.g.\ \cite{Mukhanov:2005sc}. 

A general feature of UDM models  is the possible appearance of an effective sound speed, which may become  significantly different from zero during the Universe evolution, then corresponding in general to the appearance of a Jeans length (i.e.\ a sound horizon) below which the dark fluid does not cluster (e.g.\ see \cite{Hu:1998kj, Bertacca:2007cv, Pietrobon:2008js}). Moreover, the presence of a non-negligible speed of sound can modify the evolution of the gravitational potential, producing a strong Integrated Sachs Wolfe (ISW) effect \cite{Bertacca:2007cv}. Therefore, in UDM models it is crucial to study the evolution of the effective speed of sound and that of the Jeans length.
In other words, one would conclude that any UDM model should satisfy the condition  that   $k_{\rm J}^2 \gg k^2$ for all scales of cosmological interest, in turn giving an evolution for the gravitational potential $\Phi$ as  in Eq.\ (\ref{uklesskJ}):
\begin{equation}\label{Phi_k-eta<1}
 \Phi_{\rm k} \simeq A_{\rm k}\left(1 - \frac{H}{a}\int a^2 d\eta\right)\;,
\end{equation}
where $A_{\rm k} = \Phi_{\rm k}\left(0\right)T_{\rm m}\left(k\right)$, $\Phi_{\rm k}\left(0\right)$ is the primordial gravitational potential at large scales, set during inflation,  and $T_{\rm m}\left(k\right)$ is the matter transfer function, see e.g.\ \cite{Dodelson:2003ft}.

Therefore the speed of sound plays a major role in the evolution of the scalar field 
perturbations and in the growth of the over-densities. If $c_s $ is significantly 
different from zero it can alter the evolution of density of linear and non-linear 
perturbations \cite{Hu:1998kj}. When cs becomes large at late times, this leads to strong deviations from the 
usual ISW effect  of $\Lambda$CDM models \cite{Bertacca:2007cv}.

In the next section we will perform an analytical study of the Integrated Sachs-Wolfe (ISW) effect within the framework of Unified Dark Matter models based on a scalar field 
which aim at a unified description of dark energy and dark matter. Computing the angular power spectrum of the Cosmic Microwave Background temperature anisotropies 
we are able to isolate those contributions that can potentially lead to strong deviations from the usual ISW effect occurring in a $\Lambda$CDM universe. 
This helps to highlight the crucial role played by the sound speed in the unified dark matter models. 

\section{Analytical approach to the ISW effect}\label{ISW}

In this Section we focus on the contribution to the 
large-scale CMB anisotropies which is due to the evolution in time of the gravitational potential
from the epoch of last scattering up to now, the so called late Integrated Sachs-Wolfe (ISW) effect~\cite{Sachs:1967er}. Through an analytical approach 
we point out the crucial role of the speed of sound in the unified dark matter models in determining strong deviations from the usual standard ISW 
occurring in the $\Lambda$CDM models. Our treatment is completely general in that all the results depend only on the speed of sound of the dark 
component and thus it can be applied to a variety of models, including those which are not described by a scalar field but relies on a single perfect 
dark fluid. In the case of $\Lambda$CDM models the ISW is dictated by the background evolution, which causes the late time decay of the gravitational 
potential when the cosmological constant starts to dominate~\cite{Hu:1995em}. 
In the case of the unified models there is an important aspect to consider:
from the last scattering to the present epoch, the energy density of the Universe is dominated by a single dark fluid, and therefore the gravitational potential 
evolution is determined by the background and the perturbation evolution of just such a fluid. As a result the general trend 
is the appearance of a sound speed significantly different from zero at late times corresponding to the appearance of a Jeans length (or a sound horizon) 
under which the dark fluid does not cluster any more, causing a strong evolution in time of the gravitational potential 
(which starts to oscillate and decay) and thus a strong ISW effect. Our results 
show explicitly that the CMB temperature power spectrum $C_{\ell}$ 
for the ISW effect contains some terms depending on the speed of sound which give a high contribution
along a wide range of multipoles $\ell$. As the most straightforward way to avoid these critical terms one can 
require the sound speed to be always 
very close to zero. 
Moreover we find that such strong imprints from the ISW effect come primarily from the evolution 
of the dark component perturbations, rather than from the background expansion history.   

The ISW contribution to the CMB power spectrum is given by
\begin{equation}
\label{Cl}
\frac{2l+1}{4\pi}C_{l}^{\mathrm{ISW}}= \frac{1}{2\pi^2}\int_0^\infty 
\frac{dk}{k}
k^3  \frac{\left|\Theta_{l}^{\mathrm{ISW}}(\eta_0,k)\right|^2}{2l+1}\, ,
\end{equation}
where $\Theta_{l}^{\mathrm{ISW}}$ is the fractional temperature perturbation 
due to ISW effect 
\begin{equation}
\label{theta}
\frac{\Theta_{l}^{\mathrm{ISW}}(\eta_0,k)}{2l+1}=2\int_{\eta_*}^{\eta_0}
\Phi'(\tilde{\eta},k) j_l[k(\eta_0-\tilde{\eta})]d\tilde{\eta}\, ,
\end{equation}
with $\eta_0$ and $\eta_*$ the present and the last scattering 
conformal times respectively and $j_l$ are the spherical Bessel functions.
Let us now evaluate analytically the power spectrum~(\ref{Cl}). As a first step, following the 
same procedure of Ref.~\cite{Hu:1995em}, we notice that, 
when the acceleration of the Universe begins to be important, the 
expansion time scale $\eta_{1/2}=\eta(w=-1/2)$ sets a critical wavelength 
corresponding to $k \eta_{1/2}=1$. It is easy to see that if we consider the $\mathrm{\Lambda CDM}$ model then 
$\eta_{1/2}=\eta_\Lambda$ i.e. when $a_\Lambda/a_0=(\Omega_0/\Omega_\Lambda)
^{1/3} $ \cite{Hu:1995em}. Thus at this critical point we can break the integral (\ref{Cl}) in two parts~~\cite{Hu:1995em}
\begin{equation}
\label{Cl2}
\frac{2l+1}{4\pi}C_{l}^{\mathrm{ISW}}= \frac{1}{2\pi^2}\left[I_{\Theta_{l}}
(k \eta_{1/2} <1) + I_{\Theta_{l}}(k \eta_{1/2} > 1)\right]\, ,
\end{equation}
where
\begin{equation}
\label{I_theta_k-eta<1}
I_{\Theta_{l}}
(k \eta_{1/2} <1) \equiv \int_0^{1/\eta_{1/2}} 
\frac{dk}{k}k^3  \frac{\left|\Theta_{l}^{\mathrm{ISW}}(\eta_0,k)\right|^2}{2l+1}\, ,
\end{equation}
and
\begin{equation}
\label{I_theta_k-eta>1}
I_{\Theta_{l}}(k \eta_{1/2} > 1) \equiv \int_{1/\eta_{1/2}}^\infty 
\frac{dk}{k}
k^3  \frac{\left|\Theta_{l}^{\mathrm{ISW}}(\eta_0,k)\right|^2}{2l+1}\;.
\end{equation}
As explained in Ref.~\cite{Hu:1995em} the ISW integrals (\ref{theta}) 
takes on different forms in these two regimes 
\begin{eqnarray}
\label{theta-approx}
\frac{\Theta_{l \; \mathrm{ISW}}(\eta_0,k)}{2l+1} = 
\left\{ \begin{array}{ll}
2\Delta\Phi_k \; j_l[k(\eta_0-\eta_{1/2})] & k \eta_{1/2} \ll 1\\
2\Phi_k'(\eta_k) I_l / k & k \eta_{1/2} \gg 1
\end{array} \right.
\end{eqnarray}
where $\Delta  \Phi_k$ is the change in the potential from the matter-dominated
(for example at recombination) to the present epoch $\eta_0$ and $\eta_k \simeq \eta_0 - (l+1/2)/k $
is the conformal time when a given k-mode 
contributes maximally to the angle that this scale subtends on the sky, obtained at the peak of the Bessel function $j_\ell$. 
The first limit in Eq.~(\ref{theta-approx}) is obtained by approximating the Bessel function as a constant evaluated at the 
critical epoch $\eta_{1/2}$. Since it comes from perturbations of wavelengths 
longer than the distance a photon can travel during the the time $\eta_{1/2}$, a kick ($2\Delta\Phi_k$) to the photons is the main 
result, and it will corresponds to very low multipoles, since $\eta_{1/2}$ is very close to the present epoch $\eta_0$. It thus 
appears similar to a Sachs-Wolfe effect (or also to the early ISW contribution). The second limit in 
Eq.~(\ref{theta-approx}) is achieved by considering the strong oscillations of the Bessel functions in this regime, and thus 
evaluating the time derivative of the potentials out of the integral at the peak of the Bessel function, leaving the 
integral ~\cite{Hu:1995em} 
\begin{equation}
I_l \equiv \int_0^\infty  j_l(y) dy = \frac{\sqrt{\pi}}{2} 
\frac{\Gamma[(l+1)/2]}{\Gamma[(l+2)/2]}\;.
\end{equation}
With this procedure, replacing (\ref{theta-approx}a) in (\ref{I_theta_k-eta<1})
 and (\ref{theta-approx}b) in (\ref{I_theta_k-eta>1}) we can obtain the 
ISW contribution to the CMB anisotropies power spectrum (\ref{Cl}).

Now we have to calculate, through Eqs.~(\ref{ukggkJ})-(\ref{uklesskJ}) and 
(\ref{u-v}), the value of $\Phi(k,\eta)$ for $k \eta_{1/2} \ll 1$ and $k \eta_{1/2} \gg 1$.
As we will see that main differences (and the main difficulties) of the unified dark matter models  
with respect to the $\Lambda$CDM case will appear from the second regime of Eq.~(\ref{theta-approx}).

\subsection{Derivation of $I_{\Theta_{l}}$ for modes $k \eta_{1/2} <1$}

In the UDM models when $k \eta_{1/2} \ll 1$ then $c_s^2k^2 \ll \left|\theta''/\theta\right|$ 
is always satisfied. This is due to the fact that before the dark fluid starts to behave dominantly as a cosmological constant, 
for $\eta < \eta_{1/2}$, its sound speed generically is very close to zero in order to guarantee enough structure formation, and 
moreover the limit  $k \eta_{1/2} \ll 1$ involves very large scales (since $\eta_{1/2}$ is very close to the present epoch). 
For the standard $\Lambda$CDM model the condition is clearly satisfied.  
In this situation we can use the relation (\ref{uklesskJ}) 
and $\Phi_k$ can be expressed as in Eq.\ (\ref{Phi_k-eta<1}).
The integral in Eq.~(\ref{Phi_k-eta<1}) may be written as follows
\begin{equation}
\label{Ir}
\int_{\eta_i}^{\eta}a^2(\tilde{\eta})d\tilde{\eta}=
I_R + \int_{\eta_R}^{\eta} a^2(\tilde{\eta})d\tilde{\eta}\, ,
\end{equation}
where $I_R=\int_{\eta_i}^{\eta_R}a^2(\tilde{\eta})d\tilde{\eta}$ and 
$\eta_R$ is the conformal time at recombination. When $\eta_i < \eta < \eta_R$
the UDM models behave as dark matter~\footnote{In fact the Scherrer~\cite{Scherrer:2004au} and generalized 
Scherrer solutions \cite{Bertacca:2007ux} in the very early Universe, much before the equality epoch, 
have $c_s \neq 0$ and $w>0$. However  at these times the dark fluid contribution is  
sub-dominant with respect to the radiation energy density and thus there is no substantial effect on the following equations.}. 
In this temporal range the Universe is dominated by a mixture of ``matter'' and radiation and $I_R = \eta_* a_{eq} \left[\left(\xi_R^5/5\right)+\xi_R^4+\left(4\xi_R^3/3\right) \right]$, 
where $a_{eq}$ is the value of the scalar factor at matter-radiation equality,
$\xi=\eta/\eta_*$ and $\eta_*=(\rho_{eq} a_{eq}^2/24)^{-1/2}=\eta_{eq}/
(\sqrt{2}-1)$.  With these definitions it is easy to see that 
$a_R=a_{eq}(\xi_R^2+2\xi_R)$. Notice that Eq.~(\ref{Phi_k-eta<1}) is obtained in the case of adiabatic perturbations. 
Since we are dealing with UDM models based on a scalar field, there will always be an intrinsic non-adiabatic 
pressure (or entropic) perturbation. However for the very long wavelengths, $k\eta_{1/2}  \ll 1$ under consideration here such an 
intrinsic perturbation turns out to be negligible~\cite{Garriga:1999vw}. For adiabatic perturbations $\Phi_k(\eta_R) \cong (9/10)\Phi_k(0)$ 
\cite{Mukhanov:1990me} and accounting for the primordial power spectrum, $k^3 |\Phi_k(0)|^2=B k^{n-1}\;,$
 where $n$ is the scalar 
spectral index, we get from Eq.~(\ref{theta-approx}a)
\begin{eqnarray}
\label{I_theta_k-eta<1-2}
I_{\Theta_{l}}
(k \eta_{1/2} <1)  \approx  4(2l+1) B \int_0^{1/\eta_{1/2}} 
\frac{dk}{k} k^{n-1} j_l^2[k(\eta_0-\eta_{1/2})]  \times  \left| \frac{1}{10} -\frac{\mathcal{H}(\eta_0)}{a^2(\eta_0)}\left[ 
\int_{\eta_R}^{\eta_0} a^2(\tilde{\eta})d\tilde{\eta} \right] \right|^2 \; , \nonumber \\
\end{eqnarray}
where we have neglected $I_R$ since it gives a negligible contribution. 

A first comment is in order here. There is a vast class of  UDM models that are able to reproduce 
exactly the same background expansion history of the Universe as the $\Lambda$CDM model (at least from the recombination epoch on wards).
For such cases it is clear that the low $\ell$ contribution~(\ref{I_theta_k-eta<1-2}) to the ISW effect will be the same that is predicted by the 
$\Lambda$CDM model. This is easily explained considering that for such long wavelength perturbations the sound speed in fact plays 
no role.

\subsection{Derivation of $I_{\Theta_{l}}$ for modes $k \eta_{1/2} >1$}

As we have already mentioned in the previous section, in general a viable UDM must have a sound speed very close to zero 
for $\eta < \eta_{1/2}$ in order to behave as dark matter also at the perturbed level to form the structures we see today, 
and thus the gravitational potential will start to change in time for $\eta > \eta_{1/2}$. Therefore for the modes 
$k \eta_{1/2} >1$, in order to evaluate Eq.~(\ref{theta-approx}b) into Eq.~(\ref{I_theta_k-eta>1}) we can impose 
that $\eta_k > \eta_{1/2}$ which, from the definition of $\eta_k \simeq \eta_0 - (l+1/2)/k $, moves the lower limit of 
Eq.~(\ref{I_theta_k-eta>1}) to $(l+1/2)/(\eta_0-\eta_{1/2})$. 
Moreover we have that $\eta_{1/2}\sim \eta_0$. We can use this property
to estimate any observable at the value of $\eta_k$. Defining $\chi=\eta/\eta_{1/2}\, , $ and $\kappa=k \eta_{1/2}$, 
we have $a_k=a(\eta_k)=a(\chi_k)=a_0+ (da/d\chi) \big|_{\chi_0}\delta\chi_k =1-\eta_{1/2}\mathcal{H}_0(l+1/2)/\kappa$, 
taking $a_0=1$, and 
\begin{equation}
\label{Phi_eta-k_approx}
\frac{d\Phi_{k}}{d\chi}(\chi_k)= \eta_{1/2} \Phi'(\eta_k)=
\frac{d\Phi_{k}}{d\chi}\Bigg|_{\chi_0} - \frac{d^2\Phi_{k}}{d\chi^2}
\Bigg|_{\chi_0}\left(\frac{l+1/2}{\kappa}\right)\; ,
\end{equation}
where $\delta\chi_k=\chi_k - \chi_0=(\eta_k - \eta_0)/\eta_{1/2}=-(l+1/2)/\kappa\;.$
Notice that the expansion~(\ref{Phi_eta-k_approx}) is fully justified, since as already mentioned above, the minimum value of $\kappa$ 
in Eq.~(\ref{I_theta_k-eta>1}) moves to $(l+1/2)/(\eta_0/\eta_{1/2}- 1)$, making $\delta\chi_k$ much less than 1. 
Therefore we can write 
\begin{eqnarray}
\label{Theta-approx}
\frac{\left|\Theta_{l\;\mathrm{ISW}}(\eta_0,k)\right|^2}{(2l+1)^2}=
4 \left|\frac{\Phi_k'(\eta_k) I_l}{k}\right|^2=\frac{4I_l^2}{\kappa^2}
\left|\frac{d\Phi_k}{d\chi}(\chi_k)\right|^2
= \frac{4I_l^2}{\kappa^2}\left[\left|\frac{d\Phi_{k}}{d\chi}(\chi_0)
\right|^2\right.\nonumber \\ 
\left. -2 \frac{d\Phi_{k}}{d\chi}(\chi_0) \frac{d^2\Phi_{k}}{d\chi^2}
(\chi_0) \left(\frac{l+1/2}{\kappa}\right)  \left|\frac{d^2\Phi_{k}}{d\chi^2}
(\chi_0)\right|^2\left(\frac{l+1/2}{\kappa}\right)^2\right]\;.
\end{eqnarray}
In this case, during $\eta_{1/2}<\eta<\eta_0$, there will be perturbation modes whose wavelength stays bigger than the Jeans 
length or smaller than it, i.e. we have to consider both possibilities $c_s^2k^2 \ll \left|\theta''/\theta\right|$ and  
$c_s^2k^2 \gg \left|\theta''/\theta\right|$. In general the sound speed can vary with time, and in particular 
it might become significantly different from zero at late times. However, just as a first approximation, 
we exclude the intermediate situation because usually $\eta_{1/2}$ is very close to $\eta_0$ 
(see also Ref.~\cite{Bertacca:2007cv}).

\subsubsection{Perturbation modes on scales bigger than the Jeans length}

We can see that for $n\sim1$ and for $l \gg 1$ the contribution to the angular power spectrum from the modes under consideration is 
\[\frac{l (l+1)}{4\pi} C_l^{ISW} = l(l+1)\frac{I_{\Theta_{l}}(k \eta_{1/2} > 1)}{2\pi^2(2l+1)} \sim \frac{1}{l}\;.\] 
In other words we find a similar slope as found for the $\mathrm{\Lambda CDM}$ model in Refs.~\cite{Hu:1995em,Starobinsky}. 
Recalling the results of the previous section, this means that in UDM models 
the contribution to the ISW effect from those perturbations that are outside the Jeans length is very similar to the one produced in a 
$\Lambda$CDM model. The main difference on these scales will be present if the background evolution is different from the one in the 
$\Lambda$CDM model, but for the models where the background evolution is the same, as those proposed in 
Refs.~\cite{Scherrer:2004au,Bertacca:2007ux,Gorini:2003wa,Gorini:2005nw,Quercellini:2007ht} no difference can be observed.    

\subsubsection{Perturbation modes on scales smaller than the Jeans length}

When $c_s^2k^2 \gg \left|\theta''/\theta\right|$ one must use the solution  
(\ref{ukggkJ}) and through the relation~(\ref{u-v}a)  the gravitational potential is given by 
\begin{equation}
\label{usol}
\Phi_k(\eta)=\frac{1}{2}\left[(p+\rho)/c_s\right]^{1/2}(\eta) 
\mathit{C}_k(\eta_{1/2})
\cos\left(k\int_{\eta_{1/2}}^{\eta}c_s(\tilde{\eta})d\tilde{\eta} \right)\, .
\end{equation}
In Eq.~(\ref{usol}) $\mathit{C}_k(\eta_{1/2}) = \Phi_k(0)\mathit{C}_{1/2}$ is a constant of integration where 
\begin{equation}
\mathit{C}_{1/2}=2\frac{\left[1-\frac{\mathcal{H}(\eta_{1/2})}{a^2(\eta_{1/2})}
\left(I_R+\int_{\eta_{R}}^{\eta_{1/2}}a^2(\tilde{\eta})d\tilde{\eta}\right)
\right]}{\left[(p+\rho)/c_s\right]^{1/2}(\eta_{1/2})}\; ,
\end{equation}  
and it is obtained under the approximation that for $\eta < \eta_{1/2}$ one can use the longwavelength solution~(\ref{Phi_k-eta<1}), 
since for these epochs
the sound speed must be very close to zero. Notice that Eq.~(\ref{usol}) shows clearly that the gravitational potential is oscillating 
and decaying in time. 
Defining  $\overline{C}^2 =\mathit{C}_{1/2}^2
[(p+\rho)/c_s](\eta_0)/4$, we take the time derivative of the gravitational potential appearing in Eq.~(\ref{theta-approx}b) by employing 
the expansion of Eq.(\ref{Theta-approx}). We thus find that, for $c_s \sim 1$, Eq.~(\ref{I_theta_k-eta>1}) yields 
the potentially most dangerous term
\begin{equation}
\label{I_theta_k-eta_g_1-cs_k_gg_theta''/theta}
\frac{I_{\Theta_{l}}(k \eta_{1/2} > 1)}{2l+1} \sim 4 \overline{C}^2 
B I_l^2 \eta_{1/2}^{n-1} \left\{
4c_s^4\big|_{\chi_0}\,(l+1/2)^2\left[\int_{\frac{l+1/2}{\chi_0 - 1}} 
^\infty \frac{d\kappa}{\kappa} \kappa^{n-1} \cos^2(\mathit{D}_0 \kappa)
\right]\right\}\, ,
\end{equation}
with $\mathit{D}_0=\int_1^{\chi_0} c_s(\tilde{\chi})d\tilde{\chi}$.
Such a term makes the angular power spectrum $l(l+1)C_l$ to scale as $l^3$ 
until $l\approx 25$. This angular scale is obtained by considering the peak of the Bessel functions in correspondence of the cut-off scale 
$k_{eq}$, $l \approx k_{eq}(\eta_0-\eta_{1/2})$. In fact, for smaller scales, $l(l+1)C_l$ will decrease as $1/ \ell$.  
This is due to a natural cut-off in the various integrals which is introduced for those modes that enter the 
horizon during the radiation dominated epoch, due to the Meszaros effect that the matter fluctuations will suffer until the full matter domination epoch. 
Such a cut-off will show up in the gravitational potential and in the various integrals of  Eq.~(\ref{I_theta_k-eta_g_1-cs_k_gg_theta''/theta}) as a 
$(k_{eq}/k)^4$ factor, where $k_{eq}$ is the wavenumber of the Hubble radius at the equality epoch. 

\subsection{Discussion of some examples}

Most UDM models have several properties in common. 
It is easy to see that in Eq.~(\ref{Ir}) $I_R$  is negligible because of the low value
of $a_{eq}$.
Moreover in the various models usually we have that strong differences with respect to the ISW effect in the $\Lambda$CDM case
can be produced from perturbations on those scales that are inside the Jeans length as the photons pass through them. For these scales 
the perturbations of the UDM fluid play the main role. On larger scales instead we 
find that they play no role and ISW signatures different from the $\Lambda$CDM case can come only from the different background expansion histories.   
We have found that when $k^2 \gg k^2_J=c_s^{-2}
\left|\theta''/\theta\right|$ (see~(\ref{diff-eq_u})) one must take care of  the term in  
Eq.~(\ref{I_theta_k-eta_g_1-cs_k_gg_theta''/theta}).
Indeed this term grows faster than the other integrals contained in (\ref{I_theta_k-eta_g_1-cs_k_gg_theta''/theta})
when $l$ increases up to $l\approx 25$. It is responsible for a strong ISW effect and hence, in the CMB power spectrum 
$l(l+1)C_l/(2\pi)$, it will cause a decrease in the peak to plateau ratio (once the CMB power spectrum is normalized).  
In order to avoid this effect, a sufficient (but not necessary) condition
is that the models have satisfy the condition $c_s^2k^2 < \left|\theta''
/\theta\right|$ for the scales of interest. The maximum constraint is found in correspondence of the scale at which the contribution 
Eq.~(\ref{I_theta_k-eta_g_1-cs_k_gg_theta''/theta}) 
takes it maximum value, that is 
$k \approx k_{eq}$. For example in the 
Generalized Chaplygin Gas model (GCG), i.e when 
$p=-\Lambda^{1/(1+\alpha)}/\rho^\alpha$ and $c_s^2=-\alpha w$ (see Section \ref{pkUDM}), we deduce that 
$|\alpha| <  10^{-4}$ (see Refs.~\cite{Bento:2002ps}
\cite{Carturan:2002si} \cite{Amendola:2003bz}). This is also in agreement with 
the finding of Ref.~\cite{Sandvik:2002jz} which performs an analysis on the mass power spectrum and gravitational lensing constraints, 
thus finding a more stringent constraint. \\
As far as the generalized  Scherrer solution models \cite{Bertacca:2007ux} are concerned, in these models the pressure of the UDM 
fluid is given by $p=g_n (X-X_0)^n-\Lambda$, where $g_n$ is a suitable constant and $n>1$ (see Section \ref{pkUDM}). 
The case $n=2$ corresponds to unified model proposed by Scherrer~\cite{Scherrer:2004au}. In this case we find that imposing the 
constraint $c_s^2k^2 < \left|\theta''/\theta\right|$ for the scales of interest we get
$\epsilon=(X-X_0)/X_0 < (n-1)\;10^{-4}$.\\
If we want now to study in greater detail what happens in the GCG model
when $c_s^2k^2 \gg \left|\theta''/\theta\right|$ we discover the following 
things:
\begin{itemize}
\item
for $10^{-4}<\alpha \leq 5 \times 10^{-3}$, where  we
are in the ``Intermediate case''. Now $c_s^2=-\alpha w$ is very small
and the background of the cosmic expansion history of the 
Universe is very similar to the $\mathrm{\Lambda CDM}$ model.
In this situation the pathologies, described before, 
are completely negligible.
\item
For $ 6 \times 10^{-3} < \alpha \leq 1$ a very strong ISW effect is produced; one estimates the same order of 
magnitude for the decrease of the peak to plateau ratio in the anisotropy spectrum  
$l(l+1)C_l/(2\pi)$ (once it is normalized) obtained numerically in Ref.~\cite{Carturan:2002si} 
(having assumed that the production of the peaks during the acoustic oscillations at
recombination is similar to what happens in a $\Lambda$CDM model, since at recombination 
the effects of the sound speed are negligible).
\end{itemize}
An important observation arises when considering those UDM models
that reproduce the same cosmic expansion history of the 
Universe as the $\mathrm{\Lambda CDM}$ model. Among these models one can impose the condition  
$w=-c_s^2$ which, for example, is predicted by UDM models with a kinetic term of Born-Infeld type 
\cite{Bertacca:2007ux, Gorini:2003wa, Gorini:2005nw, Padmanabhan:2002sh}. In this case, computing the integral 
in Eq.~(\ref{I_theta_k-eta_g_1-cs_k_gg_theta''/theta}) 
which gives the main contribution to the ISW effect 
one can estimate that the corresponding decrease of peak to plateau ratio is about
one third with respect to what we have in the GCG when the value of $\alpha$ 
is equal to 1. The special case $\alpha=1$ is called  ``Chaplygin Gas'' 
(see for example \cite{Bilic:2001cg}) and it is characterized by a background equation of state $w$ which evolves in a different way 
to the standard $\Lambda$CDM case. \
From these considerations we deduce that this specific effect stems
only in part from  the background of the cosmic expansion history of the 
Universe and that the most relevant contribution to the ISW effect is due to the value of the speed of sound $c_s^2$.

Let us now make some comments about  a particular class of the generalized Chaplygin gas models where the sound speed can be larger 
than the speed of light at late times, i.e. when $\alpha>1$ (see, for example, \cite{Gorini:2007ta, Urakawa:2009jb, Piattella:2009da}).
In particular, in \cite{Piattella:2009da}, the author  finds that the new constraint $\alpha > 350$. Indeed, for this range of values, the Jeans 
wavenumber is sufficiently large that the resulting ISW effect is not strong. In this case the Chaplygin gas is characterised by a fast transition \cite{Piattella:2009kt}.
However this particular model is ruled out because the transition from a pure CDM-like early phase to a post-transition 
$\Lambda$CDM-like late epoch is nearly today ($z \sim 0.22$). In fact, as discussed in Ref.~\cite{Piattella:2009kt} and in Section 
\ref{p-rho_plane}, the fast transition has to take place sufficiently far in the past. Otherwise, we expect that it would be problematic 
to reproduce the current observations related to the UDM parameter $w$, for instance it would be hard to have a good fit of the CMB and matter power spectra. 

\section{Purely kinetic Lagrangians}\label{pkUDM}

In this section we focus mainly on 
Lagrangians $\mathcal{L}$ (i.e. the pressure $p$) that depend only on\footnote{This section is largely based on Ref.\ \cite{Bertacca:2007ux}.}  $X$.
Defining $p(\rho)=g(X)$, we have to solve the equation 
\begin{equation}
\label{p_X}
\rho(g(X))=2X \frac{\partial g(X)}{\partial X} - g(X)
\end{equation}
when $X$ is time-like. 
Then, from Eq.~(\ref{eq_phi}) we get 
\begin{equation}
\label{background_lambda=0}
\left(\frac{\partial g}{\partial X}+
2X\frac{\partial^2 g}{\partial X^2} \right)\frac{dX} {dN} 
+ 3 \left(2 X \frac{\partial g}{\partial X}\right) = 0\;,
\end{equation}
where $N=\ln a$.
We can immediately note that a purely kinetic Lagrangian,  through Eq.~(\ref{p_X}), 
(see for example Ref.~\cite{Bertacca:2007ux}), can be described as a perfect fluid 
whose pressure $p$ is uniquely determined by the energy density, since both depend on a single degree of freedom, the 
kinetc term $X$.  In this case $c_s^2=p'/\rho'$ corresponds to the usual adiabatic sound speed. 
Obviously if we consider ${\it a~priori}$  a barotropic or adiabatic
equation of state, $p=p(\rho)$ can be described through a purely
kinetic k-\emph{essence} Lagrangian, if the inverse function of 
$\rho=\rho(p)$ exists. In Section \ref{p-rho_plane}, we will use the pressure-density plane
to analyze the properties that a general barotropic UDM model has to fulfil in order to be viable (see also  \cite{Piattella:2009kt}).

Now we want to make a general study of the attractor solutions in this case.
From Eq.~(\ref{eq_phi}) (see Ref.\ \cite{Bertacca:2007ux}) we obtain the following nodes,
\begin{equation}
\label{}
1) \quad \quad
X=\hat{X} = 0  \;, \quad \quad \quad \quad \quad 
2) \quad \quad 
\frac{d g}{d X}\biggl|_{\hat{X}} = 0 \;,  
\end{equation}
with $\hat{X}$ a constant. 
Both cases correspond to $w=-1$, as one can read from Eq.~(\ref{w}). 

In these cases we have either $X = 0$ or $\partial g/\partial X = 0$ on the 
node. 
We know from Eq.~(\ref{background_lambda=0}) that $X$ can only decrease in time 
down to its minimum value. This implies that $w$, from Eq.~(\ref{w}), will tend to 
$-1$ for $N \rightarrow \infty$. 

At this point we can study the general solution of the differential equation 
(\ref{background_lambda=0}). For $X \neq 0$ and 
$\partial g/\partial X  \neq 0$ the solution is \cite{Scherrer:2004au} 
\begin{equation}
\label{sol_back}
X\left(\frac{\partial g}{\partial X}\right)^{2} =  k a^{-6}
\end{equation}
with $k$ a positive constant. 
This solution has been also derived, although in a different form, in Ref.~\cite{Chimento:2003ta}.  
As $N \rightarrow \infty$, $X$ or $d g/d X$  
(or both) must tend to zero, which shows that, depending on the specific 
form of the function $g(X)$, each particular solution will converge 
toward one of the nodes above.
From Eq.~(\ref{sol_back}), for $N \rightarrow \infty$, 
the value of $X$ or $\partial g/\partial X$  
(or of both of them) must tend to zero. Then, it is immediate to conclude 
that $w \rightarrow -1$ is an attractor for $N \rightarrow \infty$
and confirms that each of the above solutions will be an attractor 
depending on the specific form of the function $g(X)$.  

In what follows we will provide some examples of stable node solutions of
the equation of motion, some of which have been already studied in the 
literature. The models below are classified on the basis of the stable node to 
which they asymptotically converge. 

\subsection{Case 1): Generalized Chaplygin gas}

An example of case 1) is provided by the Generalized Chaplygin (GC) 
model (see e.g. Refs.~\cite{Kamenshchik:2001cp,Bilic:2001cg,Bento:2002ps,Carturan:2002si,Amendola:2003bz,Makler:2003iw,Bean:2003ae,Sandvik:2002jz})
whose equation of state has the form
\begin{equation}
\label{p_GC}
p_{GC}= -\rho_* \left(\frac{\rho_{GC}}{-p_*}\right)^{\frac{1}{\gamma}} \;,
\end{equation}
where now $p_{GC}=p$ and $\rho_{GC}=\rho$ and 
$\rho_*$ and $p_*$ are suitable constants. 

Plugging the equation of state (\ref{p_GC}) into the the continuity equation
$ d \rho_{GC}/dN + 3 (\rho_{GC} + p_{GC}) = 0$, we can write $p_{GC}$ and $\rho_{GC}$ as function of $a$.
Indeed
\begin{equation}
\label{}
p_{GC} = -\left(\frac{-p_*}{\rho_{*}^{\gamma}}\right)^{1/(1-\gamma)} 
\left[1 + \nu a^{3\frac{1-\gamma}{\gamma}} \right]^{-\frac{1}{1-\gamma}} 
\end{equation}
\begin{equation}
\label{}
\rho_{GC} = \left(\frac{-p_*}{\rho_{*}^{\gamma}}\right)^{1/(1-\gamma)} 
\left[1 + \nu a^{3\frac{1-\gamma}{\gamma}} \right]^{-\frac{\gamma}{1-\gamma}} 
\end{equation}
with $\nu={\rm const.~}$. We note that, when $a$ is small, we have $\rho_{GC} \propto a^{-3}$. In other words, 
this model behaves as DM. Meanwhile, in the late epoch (i.e. $a \gg 1$), it behaves as a cosmological constant. 

Instead, through Eq.\ (\ref{p_X}),
we can obtain the pressure and the energy density as functions 
of $X$. Then
\begin{equation}
\label{}
g(X)= -\left(\frac{-p_*}{\rho_{*}^{\gamma}}\right)^{1/(1-\gamma)} 
\left[1 - \mu X^{\frac{1-\gamma}{2}} \right]^{\frac{1}{1-\gamma}}
\end{equation}
\begin{equation}
\label{}
\rho_{GC} = \left(\frac{-p_*}{\rho_{*}^{\gamma}}\right)^{1/(1-\gamma)} 
\left[1 - \mu X^{\frac{1-\gamma}{2}} \right]^{\frac{\gamma}{1-\gamma}}
\end{equation}
where $\mu$ is a constant. To connect $\mu$ and $\nu$ we have 
to use Eq.~(\ref{sol_back}). We get 
\begin{equation}
\nu = \mu^{\gamma} \left(\frac{1}{4 k}\right)^{\frac{1-\gamma}{2\gamma}} 
\left(\frac{-p_*}{\rho_{*}^{\gamma}}\right)^{\gamma}.
\end{equation}

Since $c_s^2=w/\gamma$, it is necessary for our scopes to consider the case 
$\gamma < 0$, so that $c_s^2 >0$.
Note that $\gamma = - 1$ corresponds to the standard ``Chaplygin gas'' model. 
Let us obviously consider $\mu > 0$ and $\nu > 0$.

Let us conclude this section mentioning two more models that fall into this class of solution.
The first was proposed in Ref.~\cite{Chimento:2003zf}, in which $g=b\sqrt{2X} - \Lambda$ (with $b$ a 
suitable constant) satisfying the constraint $p=-\Lambda$ along the attractor solution 
$X_0=0$. This model, however is well-known to imply a diverging speed of 
sound. The second was proposed in Refs.\  \cite{Ananda:2005xp, Quercellini:2007ht, Balbi:2007mz,Pietrobon:2008js} where
the single dark perfect fluid with ``affine'' 2-parameter barotropic equation of state $p = -\Lambda + \alpha \rho$ which satisfies the constraint that $p=-\Lambda$ 
along the attractor solution $X_0=0$. For this model, we have $c_s^2=\alpha$, i.e. the speed of sound is always a constant.  The evolution of $\rho$ leading 
to $\rho(a)=\Lambda+\rho_{m0} a^{-3(1+\alpha)}\;,$ where today $\rho_{m0}=\rho(a=1)-\Lambda$.
When the pressure and the energy density are considered as functions 
of $X$ we have
\begin{equation}
\label{pidichi2}
g(X)=-\Lambda+cX^{\frac{1+\alpha}{2\alpha}};\qquad \rho=\Lambda+\frac{c}{\alpha}X^{\frac{1+\alpha}{2\alpha}},
\end{equation}
where $c=\rho_{m0}\alpha/\hat{X}^{(1+\alpha)/(2\alpha)}$ is the integration constant derived imposing the value of the fluid energy density at present and $\hat{X}$ 
is $X$ at present time. From the matter power spectrum constraints \cite{Pietrobon:2008js}, it turns out that $\alpha \lesssim 10^{-7}$.

\subsection{Case 2): Scherrer solution}

For the solution of case 1) we want to study 
the function $g$ around some $X=\hat{X} \neq 0$. 
In this case we can approximate
$g$ as a parabola with $\frac{\partial g}{\partial X}\mid_{\hat{X}} = 0$
\begin{equation}
\label{g}
g = g_{0}+g_{2}(X-\hat{X})^{2}.
\end{equation}
with $g_0$ and $g_2$ suitable constants. This solution, with 
$g_0<0$ and $g_2>0$, coincides with the model studied by
Scherrer in Ref.~\cite{Scherrer:2004au} (see also Refs.\ \cite{Giannakis:2005kr, Chimento:2004jm}). 

It is immediate to see that for $X \rightarrow \hat{X} \neq \infty$ 
and $N \rightarrow \infty$ the value of $dX/dN$ goes to zero. 
Replacing this solution into Eq.~(\ref{sol_back}) we obtain 
\begin{equation}
\label{sol_back2}
4g_{2}^{2} X (X-\hat{X})^{2} = k a^{-6} \;, 
\end{equation}
while the energy density $\rho$ becomes
\begin{equation}
\label{rho_back}
\rho = - g_{0} + 4 g_{2} \hat{X} (X-\hat{X}) + 3 g_{2} (X-\hat{X})^{2}\;. 
\end{equation}
Now if we impose that today $X$ is close to $\hat{X}$ so that 
\begin{equation}
\label{approximation_back}
\epsilon \equiv \frac{X-\hat{X}} {\hat{X}} \ll 1
\end{equation}
then Eq.~(\ref{sol_back2}) reduces to 
\begin{equation}
\label{eqX_back-approx}
X = \hat{X} \left[1 +  \left(\frac{a} {a_1}\right)^{-3}\right]
\end{equation}
with $a_1 \ll a$ and with $(1/a_1)^{-3}=[1/(2 g_{2})](k/\hat{X}^{3})^{1/2}$   for $ \epsilon \ll 1\;.$
As a consequence, the energy density becomes
\begin{equation}
\label{rho_back-approx}
\rho = -g_0 + 4g_2 \hat{X}^2  \left(\frac{a} {a_1}\right)^{-3}.
\end{equation}
In order for the density to be positive at late times, we need to  
impose $g_0 < 0$. 
In this case the speed of sound (\ref{cs}) turns out to be 
\begin{equation}
c_s^2 =  \frac{(X-\hat{X})}{(3X - \hat{X})} = \frac{1}{2} 
\left(\frac{a} {a_1}\right)^{-3} \;, 
\end{equation}
We notice also that, for  $\left(a / a_1 \right)^{-3}\ll 1$ 
we have $c_s^2 \ll 1$ for the entire range of validity of this solution.
Thus, Eq.~(\ref{rho_back-approx}) tells us that our k-essence behaves like a fluid 
with very low sound-speed with a background energy density that can be written as 
\begin{equation}
\label{rho_lambda_dm}
\rho = \rho_\Lambda + \rho_{\rm DM} \;, 
\end{equation}
where $\rho_\Lambda$ behaves like a ``dark energy'' component 
($\rho_\Lambda = {\rm const.}$) and 
$\rho_{\rm DM}$ behaves like a ``dark matter'' component 
($\rho_{\rm DM}\propto a^{-3}$).  
Note that, from Eq.~(\ref{rho_back-approx}), $\hat{X}$ must be 
different from zero in order for the matter term to be there. 
(For this particular case the Hubble parameter $H$ is a function only 
of this fluid $H^2 = \rho/3$).\\

If the Lagrangian is strictly quadratic in $X$ we can obtain explicit 
expressions for the pressure $p$ and the speed of sound $c_s$ 
in terms of $\rho$, namely 
\begin{equation}
\label{}
p = \frac{4}{3} g_0 + \frac{8}{9} g_2 \hat{X}^2 
\left\{ 1 - \left[1 + \frac{3}{4} \frac{(g_0 + \rho)}{g_2 \hat{X}^2} 
\right]^{\frac{1}{2}} \right\} + \frac{1}{3}\rho \;, 
\end{equation}
\begin{equation}
\label{}
c_s^2 = -\frac{1}{3}\left[1 + \frac{3}{4} \frac{(g_0 + 
\rho)}{g_2 \hat{X}^2} \right]^{- \frac{1}{2}} + \frac{1}{3} \;.
\end{equation}

Looking at these equations, we observe that in the early Universe
($X \gg \hat{X}$ i.e. $\rho \gg (-g_0)$) the k-\emph{essence} 
behaves like radiation. 
Therefore, the $k$-essence in this case behaves like a
low sound-speed fluid with an energy density which evolves like the sum of a
``dark matter" (DM) component with $\rho \propto a^{-3}$ and a ``dark energy" 
(DE) component with $\rho = \rm const.$.  The only difference with respect to the standard 
$\Lambda$CDM model is that in this $k$-essence model, the dark energy
component has $c_s^2 \ll 1$.
Starting from the observational constraints on $\rho_{\rm DM}$ and 
$\rho_{\rm DE}$, the value of $a_1$ is determined by the fact that 
the $k$-essence must begin to behave like dark matter prior to the epoch of 
matter-radiation equality.
Therefore, $a_1 < a_{eq}$, where $a_{eq}$ is the scale factor at the epoch of
equal matter and radiation, given by
$a_{eq}= 3 \times 10^{-4}$ (where we have imposed that the value of the scale 
factor today is $a_0=1$).  At the present time,
the component of $\rho$ corresponding to dark energy in equation 
(\ref{rho_back-approx})
must be roughly twice the component corresponding to dark matter, so
$-g_0 = 8g_2 \hat{X}^2  (1/a_1)^{-3}.$
Substituting $a_1 < a_{eq}$ into this equation, we get \cite{Scherrer:2004au}
\begin{eqnarray}
\label{limit}
\epsilon_0 =\epsilon(a_0=1) = \frac{-g_0} {g_2 \hat{X}^2} < 8 a_{eq}^3\ll 2 \times 10^{-10}.
\end{eqnarray}

In practice, if we assume that $g(X)$ has a local minimum that can be expanded 
as a quadratic form and when Eq.~(\ref{approximation_back}) 
is not satisfied (i.e. for $a < a_1$), we cannot say anything about the evolution 
of $X$ and $\rho$. The stronger bound $\epsilon_0 \leq 10^{-18}$ 
is obtained by Giannakis and Hu \cite{Giannakis:2005kr}, who considered the 
small-scale constraint that enough low-mass dark matter halos are produced 
to reionize the Universe. On the other hand the sound
speed can be made arbitrarily small during the epoch of structure formation by decreasing the
value of $\epsilon$. One should also consider the usual constraint imposed by primordial 
nucleosynthesis on extra radiation degrees of freedom, which however leads 
to a weaker constraint. Moreover the Scherrer model differs from $\Lambda$CDM in the structure of 
dark matter halos both because of  the fact that it behaves as a nearly 
pressure-less fluid instead of a set of collisioness particles. Analytically 
we will discuss this problem when we will study the static configuration 
of the UDM models, see  Section \ref{HUDM} or Ref.~\cite{Bertacca:2007fc}. 
Practically, we will see that when $X<0$, the energy density 
of the Scherrer model is negative. Thus, $p$ and $\rho$ must depend strongly on 
time. In other words, this model will behave necessarily like a fluid and, consequently, there is the 
strong possibility that it can lead to shocks in the non-linear regime \cite{Giannakis:2005kr}. 

\subsection{Case 2): Generalized Scherrer solution}\label{genScherrer}

Starting from the condition that we are near the attractor 
$X=\widehat{X}\neq 0$, we can generalize the definition of $g$, extending 
the Scherrer model in the following way
\begin{equation}
\label{p_k-n}
p = g = g_0 + g_n (X - \hat{X})^n
\end{equation}
with $n \geq 2$ and $g_0$ and $g_n$ suitable constants.\\
The density reads 
\begin{equation}
\label{rho_n}
\rho = (2n-1) g_n (X-\hat{X})^n + 2 \hat{X} n g_n (X-\hat{X})^{n-1} - g_0
\end{equation}
If $\epsilon^n = [(X-\hat{X})/\hat{X}]^n \ll 1\;,$
Eq.~(\ref{sol_back}) reduces to
\begin{equation}
\label{epsilon_n-1}
X = \hat{X} \left[1 +  \left(\frac{a} 
{a_{n-1}}\right)^{-3/(n-1)} \right]
\end{equation}
(where $a_{n-1} \ll a$) and so $\rho$ becomes
\begin{equation}
\label{rho_n-epsilon}
\rho \simeq  2 n \hat{X}^n  g_n  \left(\frac{a} 
{a_{n-1}}\right)^{-3}- g_0
\end{equation}
with $(1/a_{n-1})^{-3} = [1/(n g_n)](k / \hat{X}^{2n-1})^{1/2} $for $\epsilon^n \ll 1$.
We have therefore obtained the important result that this attractor 
leads exactly to the same terms found in the 
purely kinetic model of Ref.~\cite{Scherrer:2004au}, i.e. a cosmological constant 
and a matter term. 
One can therefore extend the constraint of Ref.~\cite{Scherrer:2004au} 
to this case, obtaining $(\epsilon_0)^{n-1} = - g_0 / (4 n \hat{X}^n  g_n) \leq 10^{-10}\;.$
A stronger constraint would clearly also 
apply to our model by considering the small-scale constraint 
imposed by the Universe reionization, as in Ref.~\cite{Giannakis:2005kr}.
If we write the general expressions for $w$ and $c_s^2$ 
we have 
\begin{equation}
\label{w_n}
w = - \biggl[1 + \left(\frac{g_n}{g_0}\right) (X-\hat{X})^n \biggr]
\biggl[ 1 - 2 n \hat{X} \left(\frac{g_n}{g_0}\right) (X-\hat{X})^{n-1}  
- (2n-1) \left(\frac{g_n}{g_0}\right) (X-\hat{X})^n \biggr]^{-1}
\end{equation}
\begin{equation}
\label{c2_n}
c_s^2 = \frac{(X-\hat{X})}{2 (n-1) \hat{X} + (2n-1) (X-\hat{X})}.
\end{equation}
For $\epsilon \ll 1$ one obtains a result similar to that of 
Ref.~\cite{Scherrer:2004au}, namely
\begin{equation}
\label{w_n-epsilon}
w \simeq -1 + 2 n \left(\frac{g_n}{ \mid g_0 \mid}\right) 
 \left(\frac{a} {a_{n-1}}\right)^{-3} \;, 
\end{equation}
\begin{equation}
\label{c2_n-epsilon}
c_s^2 \simeq \frac{1}{2(n-1)} \epsilon \;. 
\end{equation}
On the contrary, when $X \gg \hat{X}$ we obtain 
\begin{equation}
\label{w_n-big}
w \simeq c_s^2 \simeq  \frac{1}{2n-1}
\end{equation}
In this case one can impose a bound on $n$ so that at early times and/or at 
high density the k-\emph{essence} evolves like dark matter. 
In other words, when $n\gg 1$, unlike the purely kinetic case of 
Ref.~\cite{Scherrer:2004au}, the model is well behaved also at high densities. 

In the section \ref{Static_genScherrer} we study spherical collapse for the generalized Scherrer solution models.


\subsection{Studying purely kinetic models in the pressure-density plane.}
\label{p-rho_plane}

\begin{figure}[htbp]
\begin{center}
\includegraphics[width=0.6\columnwidth]{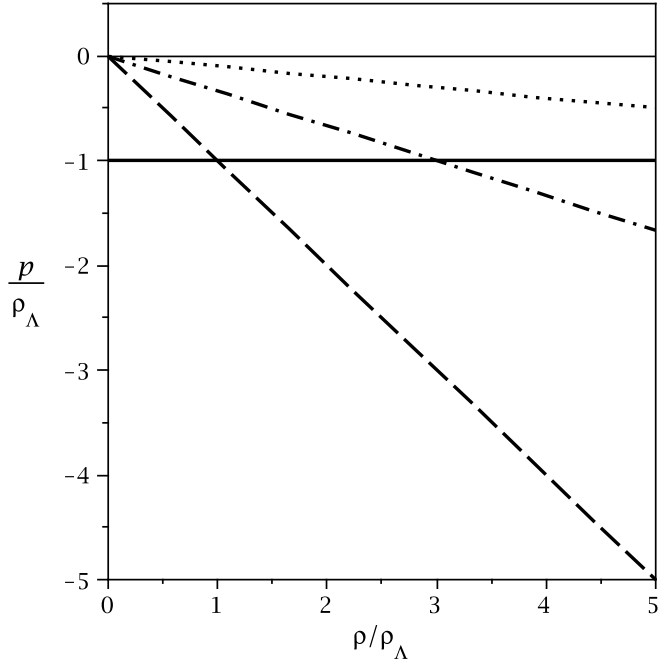}
\caption{The UDM  $p-\rho$ plane with the most important areas, (see Ref.\ \cite{Piattella:2009kt}). The dashed line represents the $p = -\rho$ line; 
the dash-dotted line represents the $p = -\rho/3$ line, the boundary between the decelerated expansion phase of the Universe and the accelerated 
one; the dotted line  $p = -\rho/10$ represents a fictitious boundary, above which  the CDM-like behaviour of the UDM fluid dominates. The pressure 
and the energy density are normalised to $\rho_\Lambda$ (where $\rho_\Lambda=\Lambda$). The $\Lambda$CDM model is represented here by the solid horizontal 
line $p/\rho_\Lambda=-1$, while the line $p=0$ represents an EdS model, i.e. pure CDM.}
\label{fig:p_rho}
\end{center}
\end{figure}
In this subsection we report some results of Ref.\  \cite{Piattella:2009kt}. Noting that purely kinetic models can be described as adiabatic single fluid  
$p=p(\rho)$, for these Lagrangians it is natural to give a graphical 
description on the $p-\rho$ plane, see  Fig.\ \ref{fig:p_rho} . Indeed, this plane gives an idea of the cosmological 
evolution of the dark fluid. Indeed, in an expanding Universe ($H>0$) Eq.\  (\ref{background_lambda=0}) implies $\dot{\rho}<0$ for a fluid satisfying 
the null energy condition $w> -1$ during its evolution, hence there exists a one-to-one correspondence between (increasing) time and (decreasing) energy density. 
Finally, in the adiabatic case  the effective speed of sound we  have introduced in Eq. (\ref{cs}) can be  written as $c_{\rm s}^2 = dp/d\rho$, therefore 
it has an immediate geometric meaning on the $p-\rho$ plane as the slope of the curve describing the EoS $p=p(\rho)$. 

For a fluid, it is quite natural to assume   $c_{\rm s}^2\geq 0$, which then  implies that the function $p(\rho)$ is monotonic, and as such it reaches the $p = -\rho$ 
line at some point $\Lambda$.\footnote{Obviously, we are assuming that during the evolution the EoS allows $p$ to  become negative, actually violating  
the strong energy condition, i.e.\  $p< -\rho/3$ at least for some $\rho>0$, otherwise the fluid would never be able to produce an accelerated expansion.} 
From the point of view of the dynamics  this is a crucial fact, because it implies  the existence of an attracting 
fixed point ($\dot{\rho}=0$) for the conservation equation (\ref{background_lambda=0}) of our UDM fluid, i.e.\  $\Lambda$ plays 
the role of an unavoidable effective cosmological constant. The Universe necessarily evolves toward  an asymptotic de-Sitter phase, a sort 
of cosmic no-hair theorem  (see  \cite{Bruni:2001pc, Bruni:1994cv} and refs.\ therein and \cite{Ananda:2005xp,Ananda:2006gf,Balbi:2007mz}).. 

We now summarise, starting from Eqs.\ (\ref{eq_u1}), (\ref{eq_u2}) and (\ref{background_lambda=0}) and taking also into account the 
current observational constraints and theoretical understanding, a list of the fundamental properties that an adiabatic UDM model has to 
satisfy in order to be viable. We then  translate these properties on the $p-\rho$ plane, see Fig.\ \ref{fig:p_rho}.

\begin{enumerate}

\item We assume the UDM to satisfy the weak energy condition: $\rho \geqslant 0$; therefore, we are only  interested in the positive half plane. In addition, 
we assume that the null energy condition is satisfied: $\rho+p\geq 0$, i.e.\ our UDM is a standard (non-phantom) fluid.
Finally, we assume that our UDM models admit a cosmological constant solution $\\Lambda$ at late time, so that an asymptotic equation of state $w = -1$ is built in.

\item We demand a dust-like behaviour back in the past, at high energies,  i.e.\ a negligible pressure $p \ll \rho$ for 
$\rho \gg \Lambda$.\footnote{Note that we could have $p\simeq -\Lambda$ and yet, if $\rho \gg \Lambda$, the Universe would still be 
in a matter-like era.}  In particular, for an adiabatic fluid we require that at recombination $|w_{\rm rec}| \lesssim 10^{-6}$, 
see \cite{Muller:2004yb,Pietrobon:2008js,Balbi:2007mz,Quercellini:2007ht}.

\item Let us consider a Taylor expansion of the  UDM EoS  $p(\rho)$ about the present energy density $\rho_0$: 
\begin{equation}
\label{affine}
p \simeq p_0 + \alpha(\rho - \rho_0)\;,
\end{equation}
i.e.\ an ``affine'' EoS model \cite{Pietrobon:2008js,Balbi:2007mz,Quercellini:2007ht,Ananda:2005xp} where $\alpha$ is the  adiabatic 
speed of sound at the present time. Clearly, these models would be represented by straight lines in Fig.\  \ref{fig:p_rho}, with $\alpha$ the slope.
The $\Lambda$CDM model, interpreted as  UDM, corresponds to the affine model (\ref{affine}) with $\alpha = 0$ (see \cite{Ananda:2005xp} 
and \cite{Balbi:2007mz,Pietrobon:2008js}) and thus  it   is represented in Fig.\ \ref{fig:p_rho} by the horizontal line $p = -\Lambda$.
From the matter power spectrum constraints on affine models \cite{Pietrobon:2008js}, it turns out that $\alpha \lesssim 10^{-7}$. 
Note therefore that, from the UDM  perspective,  today we necessarily have $w\simeq -0.7$. 
\end{enumerate}

Few comments are in order. From the points above, one could conclude that any adiabatic UDM model, in order to be viable, necessarily 
has to degenerate into the $\Lambda$CDM model, as shown in \cite{Sandvik:2002jz} for the generalised Chaplygin gas and in 
\cite{Pietrobon:2008js} for the affine adiabatic model\footnote{From the point of view of the analysis of models in the $p-\rho$ plane of Fig.\ \ref{fig:p_rho}, 
the constraints found by Sandvik {\it et al} \cite{Sandvik:2002jz} on the generalised Chaplygin gas UDM models and by   \cite{Pietrobon:2008js} 
on the affine UDM models simply amount to say that the curves representing these models are indistinguishable from the horizontal $\Lambda$CDM line.}. 
In other words, one would conclude that any UDM model should satisfy the condition $c_{\rm s}^2 \ll 1$ at all times, so that   $k_{\rm J}^2 \gg k^2$ for 
all scales of cosmological interest, in turn giving an evolution for the gravitational potential $\Phi$ as  in Eq.\ (\ref{uklesskJ}).

On the other hand, let us write down the explicit form of the Jeans wave-number:
\begin{equation}\label{kJ2analytic}
 k_{\rm J}^{2} = \frac{3}{2}\frac{\rho}{(1 + z)^2}\frac{(1 + w)}{c_{\rm s}^2}\left|\frac{1}{2}(c_{\rm s}^2 - w) - 
 \rho\frac{dc_{\rm s}^2}{d\rho} + \frac{3(c_{\rm s}^2 - w)^2 - 2(c_{\rm s}^2 - w)}{6(1 + w)} + \frac{1}{3}\right|\;.
\end{equation}
Clearly, we can obtain  a large $k_{\rm J}^{2}$ not only when $c_{\rm s}^2 \to
0$, but also when $c_{\rm s}^2$ changes rapidly, i.e.\ when the
above expression is dominated by the $\rho\; dc_{\rm s}^2/d\rho$
term. When this term is dominating in Eq.\ (\ref{kJ2analytic}), we may say that the EoS is characterised by a fast transition.

In the paper \cite{Piattella:2009kt} the authors investigate observational constraints on UDM models with fast transition, 
introducing and discussing a  toy model. In particular, they explore which values of the parameters of such a toy 
model fit the observed CMB and matter power spectra. 

\section{UDM  Scalar Field  with canonical kinetic term}\label{standardSF}

Starting from the barotropic equation of state $p=p(\rho)$ we can 
describe the system either through a purely kinetic k-\emph{essence} 
Lagrangian, as we already explained in the last section,  or through 
a Lagrangian with canonical kinetic term, as in quintessence-like models (see  Ref.\ \cite{Bertacca:2007ux}). 

In the second case we have to solve the two differential equations
\begin{eqnarray}
X - V(\varphi) & = & p(\varphi,X) \\
X + V(\varphi) & = & \rho(\varphi,X)
\end{eqnarray}
where $X=\dot{\varphi}^2/2$ is time-like. In particular, if we assume that our
model describes a unified dark matter/dark energy fluid, we can proceed as 
follows:
starting from $\dot{\rho}=-3H(p+\rho)=-\sqrt{3 \rho}(p+\rho)$ and  
$2 X =(p+\rho)=(d \varphi/d \rho)^2 \dot{\rho}^2 $ we get
\begin{equation}
\label{phi_rho}
\varphi = \pm \frac{1}{\sqrt{3}} \int^{\rho}_{\rho_0}
\frac{ d \rho' / \sqrt{\rho'}}{(p(\rho')+\rho')^{1/2}} \; ,
\end{equation}
up to an additive constant which can be dropped without any 
loss of generality. 
Inverting the Eq.~(\ref{phi_rho}) i.e. writing $\rho=\rho(\varphi)$ 
we are able to get $V(\varphi)=[\rho(\varphi)-p(\rho(\varphi))]/2$. 
Now we require that the fluid 
has constant pressure $p=-\Lambda$, i.e. that the Lagrangian of the scalar 
field is constant along the classical trajectory corresponding to  
perfect fluid behavior.  
In other words  one arrives at an exact solution with potential 
\begin{equation}
\label{ourpot}
V(\varphi)=\frac{\Lambda}{2}\left[\cosh^2\left(\frac{\sqrt 3}{2} \varphi \right) + 
1\right] \; 
\end{equation}
see also Refs.~\cite{Gorini:2003wa,Gorini:2005nw}.
For large values of $\varphi$, 
$V(\varphi) \propto \exp(\sqrt{3}\varphi)$ (equivalently, for large values of 
$-\varphi$, $V(\varphi) \propto \exp(-\sqrt{3}\varphi)$) 
and our scalar field behaves just like a pressureless dark matter 
fluid. Indeed, this asymptotic form, in the presence of an extra
radiation component, allows to recover one of the stable nodes obtained in 
Ref.~\cite{Copeland:1997et} for quintessence fields with exponential potentials, where 
the scalar field mimics a pressureless fluid.  
Under the latter hypothesis we immediately obtain 
\begin{equation}
\varphi(\rho) = \frac{2}{\sqrt{3}} {\rm arccosh} 
\left(\rho/\Lambda\right)^{1/2} \;, 
\end{equation}
which can be inverted to give the scalar field potential of Eq.~(\ref{ourpot})
as $V(\varphi) = (\rho(\varphi) + \Lambda)/2$.
One then obtains  
\begin{equation}
\label{ournode}
\dot \varphi = - \sqrt{\Lambda} \sinh\left(\frac{\sqrt 3}{2} \varphi \right) \;,  
\end{equation}
which can be immediately integrated, to give 
\begin{equation}
\label{attractraj}
\varphi(t) = \frac{2}{\sqrt{3}} \ln \left(\frac{1+\xi}{1-\xi}\right) \;, 
\quad \quad \quad \xi\equiv 
\exp\left[-\frac{\sqrt{3\Lambda}}{2}\left(t - t_*\right)\right] \;, 
\end{equation}
for $t>t_*$, with $t_*$ such that $\varphi(t\to t_*)\to \infty$. Replacing this
solution in the expression for the energy density one can easily solve the 
the Friedmann equation for the scale-factor as a function of cosmic time,  
\begin{equation}
a(t) = a_0 \frac{\sinh^{2/3}
\left[\frac{\sqrt{3\Lambda}}{2} \left(t-t_*\right)\right]}{\sinh^{2/3}
\left[\frac{\sqrt{3\Lambda}}{2} \left(t_0-t_*\right)\right]} \;, 
\end{equation}
which coincides with the standard expression for a flat, matter plus Lambda 
model \cite{Edwards}, with $\Omega_{0\Lambda}/\Omega_{0m} = \sinh^2  
[\frac{\sqrt{3\Lambda}}{2} (t_0-t_*)]$, 
$\Omega_{0\Lambda}$ and $\Omega_{0m}$ being the cosmological constant and
matter density parameters, respectively. 

Using standard criteria (e.g. Ref.~\cite{Copeland:2006wr})  
it is immediate to verify that the above trajectory corresponds to 
a stable node even in the presence of an extra-fluid (e.g. radiation) with
equation of state $w_{\rm fluid}\equiv p_{\rm fluid}/\rho_{\rm fluid}>0$, 
where $p_{\rm fluid}$ and $\rho_{\rm fluid}$ are the fluid pressure 
and energy density, respectively. 
Along the above attractor trajectory our scalar field behaves 
precisely like a mixture of pressureless matter and cosmological constant. 
Using the expressions for the energy density and the pressure we 
immediately find, for the matter energy density 
\begin{equation}
\rho_m = \rho - \Lambda = \Lambda \sinh^2 \left(\frac{\sqrt 3}{2} \varphi \right) 
\propto a^{-3} \;.
\end{equation} 
The peculiarity of this model is that the matter component appears as a 
simple consequence of having assumed the constancy of the Lagrangian.\\
A closely related solution was found by Salopek \& Stewart \cite{Salopek:1992qy}, 
using the Hamiltonian formalism. \\
To conclude this section, let us stress that, like any scalar field 
with canonical kinetic term \cite{Bludman:2007kg, Linder:2008ya}, our UDM model predicts 
$c_s^2=1$, as it is clear from Eq.~(\ref{cs}), which inhibits 
the growth of matter inhomogeneities. 
In summary, we have obtained a ``quartessence'' model which behaves exactly 
like a mixture of dark matter and dark energy along the attractor solution, 
whose matter sector, however is unable to cluster on sub-horizon scales 
(at least as long as linear perturbations are considered). 

\section{UDM Scalar Field with non-canonical kinetic term} \label{non-standardSF}

We can summarize our findings so far by stating that purely kinetic k-\emph{essence}
cannot produce a model which {\it exactly} describes a unified fluid of 
dark matter and cosmological constant, while scalar field models with canonical kinetic 
term, while containing such an exact description, unavoidably lead to 
$c_s^2=1$, in conflict with cosmological structure formation. 
In order to find an exact UDM model with acceptable speed of sound we 
consider more general scalar field Lagrangians (see  Ref.\ \cite{Bertacca:2007ux}). 

\subsection{Lagrangians of the type 
${\mathcal L}(\varphi,X) = g(X) - V(\varphi)$}\label{g-v}

Let us consider Lagrangians with non-canonical kinetic term and 
a potential term, in the form 
\begin{equation}
\label{newlagr}
{\mathcal L}(\varphi,X) = g(X) - V(\varphi) \;. 
\end{equation}
The energy density then reads 
\begin{equation}
  \label{newenergy-density}
  \rho = 2X\frac{d g(X)}
  {d X}-g(X) + V(\varphi) \;,  
\end{equation} 
while the speed of sound keeps the form of Eq.~(\ref{cs}). 
The equation of motion for the homogeneous mode reads 
\begin{equation}
\label{new_eqmot}
\left(\frac{d g}{d X}+
2X\frac{d^2 g}{d X^2} \right)\frac{dX} {dN} 
+ 3 \left(2 X \frac{d g}{d X}\right) = -  \frac{dV}{dN} \;. 
\end{equation}

One immediately finds 
\begin{equation}
p + \rho  = 2X\frac{d g(X)}{d X} \equiv 2{\cal F}(X) \;. 
\end{equation} 

One can rewrite the equation of motion Eq.~(\ref{new_eqmot}) in the form 
\begin{equation}
\left[ 2X \frac{d {\cal F}}{d X} - {\cal F} \right]
\frac{dX} {dN} + X \left( 6{\cal F} + \frac{dV}{dN} \right) = 0 \;. 
\end{equation}
It is easy to see that this equation admits 2 nodes, namely:
\begin{itemize} 
\item[1)] $d g/d X|_{\widehat{X}} =0$ and
\item[2)]$\widehat{X}=0$.
\end{itemize}
In all cases, for $N \to \infty$, the potential 
$V$ should tend to a constant, while the kinetic term can be
written around the attractor in the form 
\begin{equation}
\label{xn}
g(X) = M^4 \left(\frac{X - \widehat{X}}{M^4}\right)^n \quad \quad \quad n \geq 2 \;, 
\end{equation}
where $M$ is a suitable mass-scale and $\widehat{X}$ a constant.   
The trivial case $g(X)=X$ obviously reduces to the one of Section 4. 
 
Following the same procedure adopted in the previous section 
we impose the constraint $p=-\Lambda$, which yields the general solution
$\rho_m=2{\cal F}(X)$. 

This allows to define $\varphi=\varphi(\rho_m)$ as a solution of the 
differential equation 
\begin{equation}
\label{master1}
\rho_m = 2{\cal F}\left[\frac{3}{2}\left(\rho_m +\Lambda\right) \rho_m^2 
\left(\frac{d \varphi}{d \rho_m}\right)^2 \right] \;.  
\end{equation}

As found in the case of k-\emph{essence}, the most interesting behavior 
corresponds to the limit of large $n$ and $\widehat{X}=0$ in Eq.~(\ref{xn}), 
for which we obtain 
\begin{equation}
\rho_m \approx \Lambda \sinh^{-2}\left[\left(\frac{3\Lambda}{8M^4} \right)
^{1/2} \varphi\right] \;,
\end{equation}
leading to 
$V(\varphi) \approx \rho_m/2n - \Lambda$, and $c_s^2 = 1/(2n-1) \approx 0$. 
The Lagrangian of this model is similar to that analyzed in 
Ref.~\cite{DiezTejedor:2006qh}.\\

\subsection{Lagrangians of the type ${\mathcal L}(\varphi,X) = f(\varphi) g(X)$}\label{fg}

Let us now consider Lagrangians with a non-canonical kinetic term 
of the form  
${\mathcal L}(\varphi,X) = f(\varphi)g(X)$  (see Ref.\ \cite{Bertacca:2007ux}). 

Imposing the constraint $p=-\Lambda$, one obtains $f(\varphi)=-\Lambda /g(X)$,
which inserted in the equation of motion yields the general solution
\begin{equation}
\label{master2}
X \frac{d \ln |g|}{dX} = - \frac{\rho_m}{2\Lambda} \;. 
\end{equation}

The latter equation, together with Eq.~(\ref{master1}) define our general
prescription to get UDM models describing both DM and cosmological 
constant-like DE. 

As an example of the general law in Eq.~(\ref{master2})
let us consider an explicit solution. Assuming that  
the kinetic term is of Born-Infeld type, as in 
Refs.~\cite{Padmanabhan:2002sh,Abramo:2003cp,Gorini:2003wa,Gorini:2005nw},  
\begin{equation}
g(X)=-\sqrt{1-2X/M^4} \;, 
\end{equation}
with $M$ a suitable mass-scale, which implies $\rho = 
f(\varphi)/\sqrt{1-2X/M^4}$, we get
\begin{equation}
\label{udmsol_X}
X(a) =\frac{M^4}{2}\frac{\bar{k} a^{-3}}{1 + \bar{k} a^{-3}} \;,
\end{equation}
where $\bar{k}=\rho_m(a_*) a_*^3/ \Lambda $ and $a_*$ is the scale-factor at 
a generic time $t_*$. 
In order to obtain an expression for $\varphi(a)$, we impose that the 
Universe is dominated by our UDM fluid, i.e. $H^2=\rho/3$. This gives 
\begin{equation}
\label{udmsol}
\varphi(a) =\frac{2 M^2}{\sqrt{3\Lambda}} 
\left\{ \arctan \left[\left(\bar{k} a^{-3}\right)^{-1/2}\right] 
- \frac{\pi}{2} \right\} \;, 
\end{equation}
which, replaced in our initial ansatz $p=-\Lambda$ allows to obtain the 
expression (see also Ref.~\cite{Gorini:2003wa,Gorini:2005nw})
\begin{equation}
f(\varphi)=\frac{\Lambda}{\left\vert\cos 
\left[ \left(\frac{3\Lambda}{4M^4}\right)^{1/2}\varphi \right] 
\right\vert} \;.
\end{equation}

If one expands $f(\varphi)$ around $\varphi= 0$,  
and $X/M^4 \ll 1$ one gets the approximate Lagrangian
\begin{equation}
{\mathcal L}  \approx \frac{\Lambda}{2M^4}\dot{\varphi}^2 - 
\Lambda \left[1 + \frac{3\Lambda}{8M^4}\varphi^2\right] \;.
\end{equation}
Note that our Lagrangian depends only on the combination $\varphi/M^2$, 
so that one is free to reabsorb a change of the mass-scale in the 
definition of the filed variable. 
Without any loss of generality we can then set $M=\Lambda^{1/4}$,  
so that the kinetic term takes the canonical form in the limit $X\ll 1$. 
We can then rewrite our Lagrangian as 
\begin{equation}
\label{udmbest}
{\mathcal L} = - \Lambda \frac{\sqrt{1-2X/\Lambda} }
{\left\vert\cos \left(\frac{\sqrt{3}}{2} \varphi \right)\right\vert } \;. 
\end{equation}

This model implies that for values of 
$\sqrt{3}\varphi \approx -\pi$ and $2X/\Lambda \approx 1$, 
\begin{equation}
\cos \left(\frac{\sqrt{3}}{2} \varphi \right)  \propto  a^{3/2}\;,
\quad \quad \quad  \sqrt{1-2X/\Lambda}  \propto  a^{-3/2} \;,
\end{equation}
the scalar field mimics a dark matter fluid. In this regime the effective 
speed of sound is $c_s^2=1-2X/\Lambda \approx 0$, as desired. 

To understand whether our scalar field model gives rise to a 
cosmologically viable UDM solution, we need to check if in a Universe 
filled with a scalar field with Lagrangian (\ref{udmbest}), plus 
a background fluid of e.g. radiation, the system displays the desired
solution where the scalar field mimics both the DM and DE components. 
Notice that the model does not contain any free parameter to specify 
the present content of the Universe. 
This implies that the relative amounts of DM and DE that characterize 
the present Universe are fully determined by the value of 
$\varphi_0\equiv \varphi(t_0)$. In other words, to reproduce the present
Universe, one has to tune the value of $f(\varphi)$ in the early Universe. 
However, a numerical analysis shows that, once the initial value of 
$\varphi$ is fixed, there is still a large basin of attraction in terms of the 
initial value of $d\varphi / dt$, which can take any value such that 
$2X/\Lambda \ll 1$. 

\begin{figure}
\begin{center}
\includegraphics[width = 4in , keepaspectratio=true]{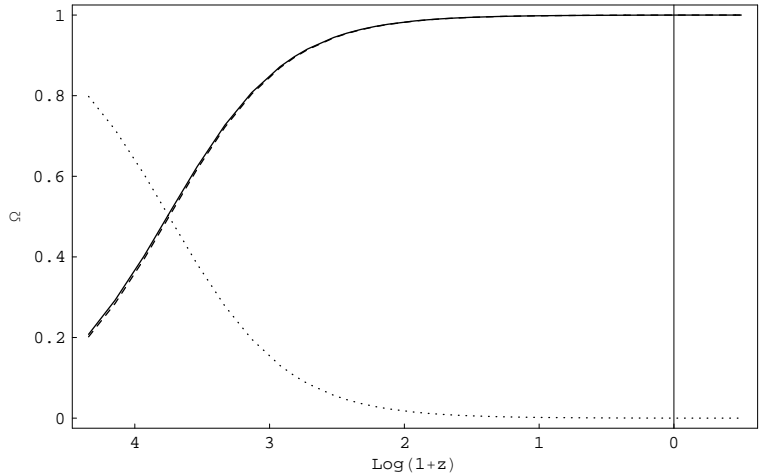}
\caption{Evolution of the scalar field density parameter vs. redshift  (see Ref.\ \cite{Bertacca:2007ux}). 
The continuous line shows the UDM density parameter; the dashed line is the 
density parameter of the DM + DE components in a standard $\Lambda$CDM model; 
the dotted line is the radiation density parameter.}
\label{fig1}
\end{center}
\end{figure}
\begin{figure}
\begin{center}
\includegraphics[width = 4in,keepaspectratio=true]{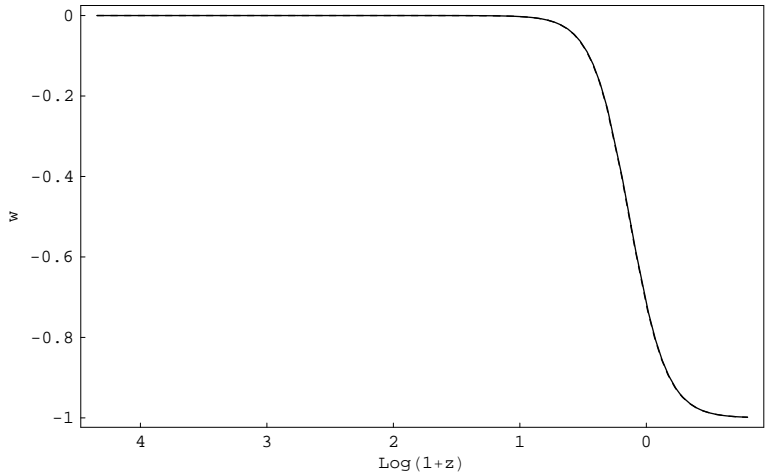}
\caption{The redshift evolution of the scalar field equation of state 
parameter $w_{\rm UDM}$ (continuous line) is compared with that of the 
sum of the DM + DE components in a standard $\Lambda$CDM model 
(dashed line), see Ref.\ \cite{Bertacca:2007ux}.}
\label{fig2}
\end{center}
\end{figure}
\begin{figure}
\begin{center}
\includegraphics[width = 4in,keepaspectratio=true]{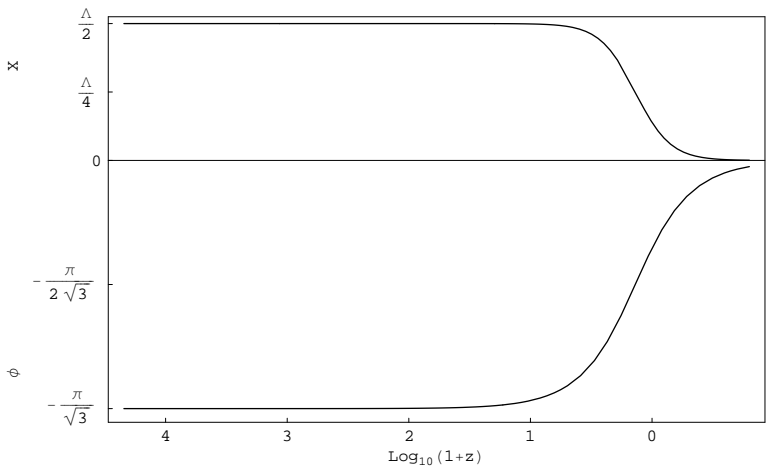}
\caption{Redshift evolution of the scalar field 
of the scalar field variables $X={\dot \varphi}^2/2$ (top) and $\varphi$
(bottom), see Ref.\ \cite{Bertacca:2007ux}.} 
\label{fig3}
\end{center}
\end{figure}

The results of a numerical integration of our system including scalar 
field and radiation are shown in Figures 2 - 4. Figure 2 shows the 
density parameter, $\Omega_{\rm UDM}$ as a function of redshift, having 
chosen the initial value of $\varphi$ so that today the 
scalar field reproduces the observed values $\Omega_{\rm DM}$ and
$\Omega_{\rm DE}$. 
Notice that the time evolution of the scalar field energy density 
is practically indistinguishable from that of a standard DM plus Lambda 
($\Lambda$CDM) model with the same relative abundances today.    
Figure 3 shows the evolution equation of state parameter $w_{\rm UDM}$; 
once again the behavior of our model is almost identical to that of 
a standard $\Lambda$CDM model for $1 + z < 10^4$. Notice that, since 
$c_s^2=-w_{\rm UDM}$, the effective speed of sound of our model
is close to zero, as long as matter dominates, as required. 
In Figure 4 we finally show the redshift evolution of the scalar field 
variables $X={\dot \varphi}^2/2$ and $\varphi$: one can easily check that
the evolution of both quantities is accurately described by the analytical 
solutions above, Eqs.~(\ref{udmsol_X}) and (\ref{udmsol}), respectively (the 
latter being obviously valid only after the epoch of matter-radiation 
equality). 

However in this model, as discussed in Ref.~\cite{Bertacca:2007cv}, 
the non-negligible value of the sound speed today gives a 
strong contribution to the ISW effect and produces an incorrect 
ratio between the first peak and the plateau of
the CMB anisotropy power-spectrum 
$l(l+1)C_l/(2\pi)$.

\section{How the Scalar Field of Unified Dark Matter Models Can Cluster}\label{recUDM}

The authors of \cite{Bertacca:2008uf} proposed a technique for constructing UDM models where the scalar field can have a 
sound speed small enough to allow for structure formation and to avoid a strong integrated Sachs-Wolfe effect in the CMB 
anisotropies which typically plague UDM models\footnote{This section is largely based on Ref.\  \cite{Bertacca:2008uf}.} 
(see also \cite{Camera:2009uz,Camera:2010wm}). In particular, they studied a class of UDM models where, 
at all cosmic times, the sound speed is small enough that cosmic structure can form.
To do so, a possible approach is to consider a scalar field
Lagrangian $\mathcal{L}$ of the form 
\begin{equation}
 \label{Lagrangian}
 \mathcal{L}= p(\varphi ,X)=f(\varphi)g(X)-V(\varphi) \; .  
\end{equation}
Therefore, by introducing the two potentials $f(\varphi)$ and $V(\varphi)$,
we want to decouple the equation of state parameter $w$ and the sound speed 
$c_s$. This condition does not occur 
when we consider either Lagrangians with purely kinetic terms or 
Lagrangians like $\mathcal{L}=g(X)-V(\varphi)$ or 
$\mathcal{L}=f(\varphi)g(X)$~(see for example \cite{Bertacca:2007ux} and the previous Sections  \ref{g-v} and \ref{fg}).
In the following subsections we will describe how 
to construct UDM models based on Eq.~(\ref{Lagrangian}), following the analysis of Ref.~\cite{Bertacca:2008uf}.

\subsection{How to construct UDM models}
\label{UDM}
Let us consider the scalar field Lagrangian of Eq. (\ref{Lagrangian}).
The energy density $\rho$, the equation of state $w$ and 
the speed of sound $c_s^2$ are
\begin{equation}
\label{rho_fg-v}
\rho(X,\varphi)=f(\varphi)\left[2X\frac{\partial g(X)}{\partial X}
-g(X)\right]-V(\varphi)\;,
\end{equation}
\begin{equation}
\label{w_fg-v}
w(X,\varphi)= \frac{f(\varphi)g(X)-V(\varphi)}{f(\varphi)
\left[2X \left(\partial g(X)/\partial X\right) -g(X)\right] -V(\varphi)}\;,
\end{equation}
\begin{equation}
\label{c_s^2_fg-v}
c_s^2(X)=\frac{\left(\partial g(X)/\partial X\right)}{\left(\partial g(X)/
\partial X\right)+ 2X\left(\partial^2 g(X)/\partial X^2\right)} \;, 
\end{equation} 
respectively.  The equation of motion (\ref{eq_phi}) becomes
\begin{equation}
 \label{eq_phi_fg-v}
\left(\frac{\partial g}{\partial X} 
+2X\frac{\partial^2 g}{\partial X^2}\right)\frac{d X}{d N}
+6X\frac{\partial g}{\partial X}+
\frac{d \ln f}{d N}\left(2X\frac{\partial g}{\partial X}
- g\right)
-\frac{1}{f}\frac{d V}{d N}=0\;.
\end{equation}

Unlike in models with a Lagrangian with
purely kinetic terms, here we have one more degree of freedom, 
the scalar field configuration itself.
This allows to impose a new condition 
to the solutions of the equation of motion.
In Ref.~\cite{Bertacca:2007ux}, the scalar field Lagrangian was required 
to be constant along the classical trajectories.
Specifically, by requiring that $\mathcal{L}=-\Lambda$ on cosmological scales, 
the background is identical to the background of 
$\Lambda$CDM.
In general this is always true. In fact, if we consider Eq.~(\ref{eq_phi})
or, equivalently, the continuity equations $(d \rho/d N)=-3(p+\rho)$,
and if we impose $p=-\Lambda$, we easily get 
\begin{equation}
\rho=\rho_{{\rm DM}}(a=1)~a^{-3}+\Lambda= \rho_{\rm DM}+ \rho_\Lambda \;,
\end{equation}
where $\rho_\Lambda$ behaves like a cosmological constant ``Dark Energy'' 
component 
($\rho_\Lambda = {\rm const.}$) and 
$\rho_{\rm DM}$ behaves like a ``Dark Matter'' component 
($\rho_{\rm DM}\propto a^{-3}$).  
This result implies that we can think the stress-energy tensor of our scalar field 
as being made of two components: one behaving like a
pressure-less fluid, and the other having negative pressure. 
In this way the integration constant $\rho_{{\rm DM}}(a=1)$
can be interpreted as the ``dark matter'' component today; consequently, 
$\Omega_m(0)=\rho_{{\rm DM}}(a=1)/(3H^2(a=1))$ and 
$\Omega_\Lambda(0)=\Lambda/(3H^2(a=1))$ are the 
density parameters of ``dark matter''and ``dark energy'' today.

Let us now describe the procedure that we will use in order to find UDM models with a small  speed of sound. 
By imposing the condition  $\mathcal{L}(X,\varphi)=-\Lambda$,
we constrain the solution of the equation of motion to live
on a particular manifold $\mathcal{M}_\Lambda$ embedded 
in the four dimensional space-time. This enables us 
to define $\varphi$ as a function of $X$ along 
the classical trajectories, i.e.
$\varphi=\mathcal{L}^{-1}(X,\Lambda)\big|_{\mathcal{M}_\Lambda}$.
Notice that therefore, by using Eq.(\ref{eq_phi_fg-v}) and 
imposing the constraint $p=-\Lambda$, i.e. 
$V(\varphi)=f(\varphi)g(X)+\Lambda$, we
can obtain the following general solution of the equation of motion
on the manifold $\mathcal{M}_\Lambda$
\begin{equation}
\label{mastereq}
2X\frac{\partial g(X)}{\partial X}f(\varphi(X))=\Lambda ~\nu ~a^{-3} \;,
\end{equation}
where $\nu\equiv \Omega_m(0)/\Omega_\Lambda(0)\,$ .
Here we have constrained the pressure to be $p=-\Lambda$.
In Section \ref{GUDM} we will describe an even more general technique 
to reconstruct
UDM models where the pressure is a free function of the scale factor $a$.

If we define the function $g(X)$, we immediately know the functional form 
of $c_s^2$ with respect to $X$ (see Eq.~(\ref{c_s^2_fg-v})). 
Therefore, if we have a Lagrangian
of the type $\mathcal{L}=f(\varphi)g(X)$ or $\mathcal{L}=g(X)-V(\varphi)$,
we are unable to decide 
the evolution of $c_s^2(X)$ along the solutions of the equation of motion
\cite{Bertacca:2007ux} because, once $g(X)$ is chosen, the 
constraint ${\cal L}=-\Lambda$ fixes immediateley the value of 
$f(\varphi)$ or $V(\varphi)$. 
On the contrary, in the case of Eq.~(\ref{Lagrangian}), we can do it 
through the function $f(\varphi(X))$. In fact, by properly defining
the value of $f(\varphi(X))$ and using Eq.(\ref{eq_phi_fg-v}), 
we are able to fix the slope of $X$ and, consequently (through $g(X)$),  
the trend of $c_s^2(X)$ as a function of the scale factor $a$.  

Finally, we want to emphasize that this approach is only a method
to reconstruct the explicit form of the Lagrangian (\ref{Lagrangian}), namely 
to separate the two variables $X$ and $\varphi$ into the
functions $g$, $f$ and $V$.

Let us now give an example where we apply this prescription.
In particular, in the following subsection, we assume a kinetic term of Born-Infeld type 
\cite{Padmanabhan:2002sh,Abramo:2003cp,Gorini:2003wa,Gorini:2005nw}.
Other examples (where we have the kinetic term $g(X)$
of the Scherrer model \cite{Scherrer:2004au} or where
we consider the generalized Scherrer solutions \cite{Bertacca:2007ux}) 
are reported in Ref.\  \cite{Bertacca:2008uf}.

\subsubsection{Lagrangians with Born-Infeld type kinetic term}
\label{BIL0}
Let us consider the following kinetic term
\begin{equation}
g(X)=-\sqrt{1-2X/M^4} \;, 
\end{equation}
with $M$ a suitable mass scale.
We get
\begin{equation}
\label{mastereq-BI}
\frac{2X/M^4}{\sqrt{1-2X/M^4}}f(\varphi(X))=\Lambda ~\nu ~a^{-3}\;,
\end{equation}
and
\begin{equation}
\label{c_s-BI}
c_s^2(X)=1-2X/M^4\;.
\end{equation}

In the next subsection, we give 
a Lagrangian where the sound speed can be small. 
It is important to emphasize that the models described here and 
in the next subsection satisfy
the weak energy conditions $\rho\ge 0$ and $p+\rho\ge 0$.

\subsection{UDM models with Born-Infeld type kinetic term and a low speed of sound}
\label{BIL}

Let us consider for $f$ the 
following definition 
\begin{equation}
\label{f-B_3}
f(\varphi(X))= \frac{\Lambda}{\mu} ~\frac{2X/M^4-h}
{2X/M^4\left(1-2X/M^4\right)^{1/2}}\;,
\end{equation}
where $h$ and $\mu$ are appropriate positive constants. 
Moreover, we impose that $h<1$.
Thus we get
\begin{equation}
\label{X-B_3}
X(a) =\frac{M^4}{2}\frac{h + \mu \nu a^{-3}}{1 + \mu \nu a^{-3}} 
\quad {\rm or} \quad 
\left(\frac{d \varphi}{d N}\right)^2 = \frac{3 M^4}{\Lambda} 
\;\frac{h + \mu \nu a^{-3}}
{\left(1 + \nu a^{-3}\right) \left(1 + \mu \nu a^{-3}\right)}\;,
\end{equation}
and, for $c_s^2$, we obtain the following relation  
\begin{equation}
c_s^2(a)=\frac{1-h }{1 + \mu \nu a^{-3}}\;.
\end{equation}
Therefore, with the definition (\ref{f-B_3}) and using 
the freedom in choosing the value of $h$, 
we can shift the value of $c_s^2$ for  $a \rightarrow \infty$.
Specifically, $h=1-c_{\infty}^2$ where 
$c_{\infty}=c_s(a \rightarrow \infty)$.
At this point, by considering the case where $h=\mu$ (which makes the 
equation analytically integrable), we can
immediately obtain the trajectory $\varphi(a)$, namely 
\begin{equation}
\label{phi-B_3}
\varphi(a) =\left(\frac{4 h M^4}{3\Lambda} \right)^{1/2}
{\rm arc}\sinh\left( \nu h a^{-3}\right)^{-1/2} \;.
\end{equation}
Finally, we obtain 
\begin{equation}
f(\varphi)=\frac{\Lambda(1-h)^{1/2}}{h}
\frac{\cosh\left[\left(\frac{3\Lambda}{4hM^4}\right)^{1/2}
\varphi\right]}{\sinh\left[\left(\frac{3\Lambda}{4hM^4}\right)^{1/2}
\varphi\right]\left\{1+h\sinh^2\left[\left(\frac{3\Lambda}{4hM^4}\right)^{1/2}
\varphi\right]\right\}} \;,
\end{equation}
and
\begin{equation}
V(\varphi) = \frac{\Lambda}{h}\frac{\left\{h^2\sinh^2
\left[\left(\frac{3\Lambda}{4hM^4}\right)^{1/2}
\varphi\right]+2h-1\right\}}{1+h\sinh^2
\left[\left(\frac{3\Lambda}{4hM^4}\right)^{1/2}
\varphi\right]} \;.
\end{equation}
This result implies that in the early universe 
$\sqrt{3\Lambda/(4hM^4)}~ \varphi \ll 1$ and 
$2X/M^4 \approx 1$, and we obtain 
\begin{eqnarray}
f(\varphi)\approx \left(\frac{4hM^4}{3\Lambda}\right)^{1/2}
\frac{\Lambda\sqrt{1-h}}{h} 
~\frac{1}{\varphi}~\propto ~a^{3/2}\;,
\quad \quad \quad \left|g(X)\right|= \sqrt{1-2X/\Lambda}
~\propto ~ a^{-3/2}\;,\nonumber \\ 
|V(\varphi)|~ \longrightarrow ~\left|\frac{\Lambda(2h-1)}{h}\right| ~\ll~  
f(\varphi)\left(2X\frac{\partial g(X)}{\partial X}-g(X)\right)
~\propto ~ a^{-3}\;.
\end{eqnarray}
In other words, we find, for $ f(\varphi)$ and $ g(X)$,
a behaviour similar to that  we have studied in Section \ref{fg}, as also in Ref.~\cite{Bertacca:2007ux}.

When $a \rightarrow \infty$, we have $\varphi \rightarrow \infty$ and
$2X/M^4 \rightarrow h$. Therefore
$$
f(\varphi) g(X) \longrightarrow 0\;, \quad \quad \quad  
V(\varphi)~ \longrightarrow \Lambda \;,
$$
that is, for $a \rightarrow \infty$, the dark fluid of this UDM model will 
converge to a Cosmological Constant.

In Ref.~\cite{Bertacca:2008uf} the authors analytically show that, 
once the initial value of 
$\varphi$ is fixed, there is still a large basin of attraction in terms of the 
initial value of $d\varphi / dt$, which can take any value such that 
$2X/M^4 \ll 1$. Moreover, Ref.~\cite{Bertacca:2008uf} investigates
the kinematic behavior of this UDM fluid during the
radiation-dominated epoch. 

\begin{figure}
\centerline{\includegraphics[width = 5in,keepaspectratio=true]{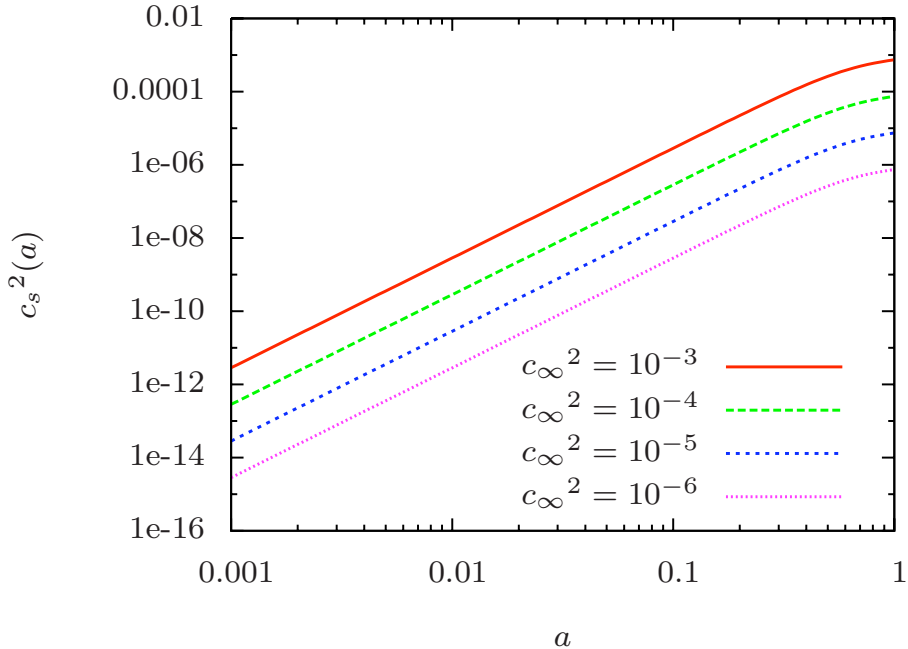}}
\caption{Sound speed ${c_s}^2(a)$ for different values of ${c_\infty}^2=10^{-6},10^{-5},10^{-4},10^{-3}$ from bottom to top (see  \cite{Camera:2009uz}).}
\label{cs2}
\end{figure}

We can conclude that, once it is constrained to yield
the same  background evolution as $\Lambda$CDM and 
we set an appropriate value of $c_{\infty}$, 
this UDM model provides a sound speed small
enough that i) the dark fluid can cluster and ii) the Integrated Sachs-Wolfe contribution 
to the CMB anisotropies is compatible with observations.
Figure \ref{cs2} shows an example of the dependence of $c^2_s$ on $a$
for different values of $c_{\infty}$. 

\begin{figure}
\centerline{\includegraphics[width = 5in,keepaspectratio=true]{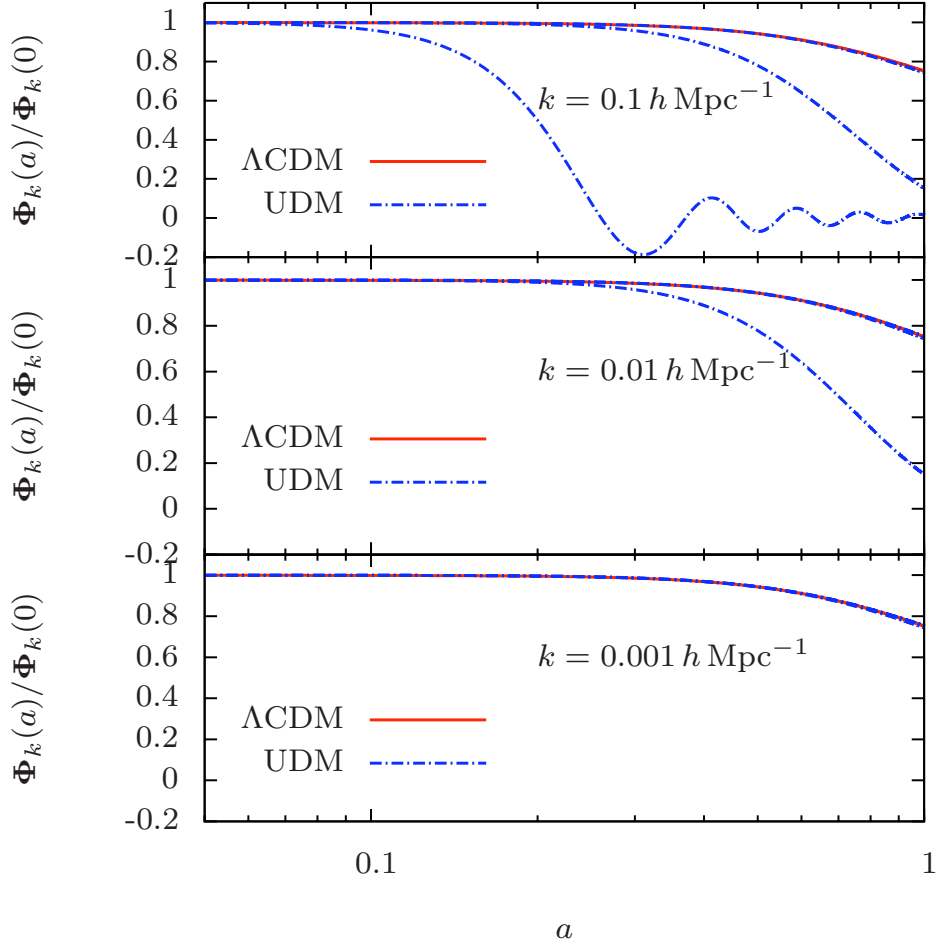}}
\caption{Normalized potentials $\Phi_k(a)/\Phi_k(0)$ are shown for $\Lambda$CDM (solid) and UDM (dot-dashed) (see \cite{Camera:2009uz}). 
The lower panel shows potentials at $k=0.001\,h\,\mathrm{Mpc}^{-1}$, the medium panel at $k=0.01\,h\,\mathrm{Mpc}^{-1}$ and the upper panel 
at $k=0.1\,h\,\mathrm{Mpc}^{-1}$. UDM curves are for ${c_\infty}^2=10^{-6},10^{-4},10^{-2}$ from top to bottom, respectively. At small $c_\infty$, 
$\Lambda$CDM and UDM curves are indistinguishable.}\label{potentials_k-cinf}
\end{figure}

In Fig.~\ref{potentials_k-cinf} we present some Fourier components $\Phi_k(a)$ of the gravitational potential, 
normalized to unity at early times (see \cite{Camera:2009uz}).
As we can note from this figure, the possible appearance of a sound speed significantly different from zero at late 
times corresponds to the appearance of a Jeans' length under which the dark fluid does not cluster any more, causing a strong evolution 
in time of the gravitational potential. By increasing the sound speed, the potential starts to decay earlier in time, oscillating around zero 
afterwards. Moreover, at small scales, if the sound speed is small enough, UDM reproduces $\Lambda$CDM. This reflects the dependence 
of the gravitational potential on the effective Jeans' length ${\lambda_J}(a)$ \cite{Bertacca:2007cv}.

Finally, in Ref.~\cite{Camera:2009uz} the authors show for this UDM model the lensing signal in linear theory as produced in $\Lambda$CDM 
and UDM; as sources, they consider the CMB and background galaxies, with different values of the peak and different shapes of their redshift 
distribution. For sound speed lower than $c_\infty=10^{-3}$, in the window of multipoles $l\gtrsim10$ (Limber's approximation) and where our 
ignorance on non--linear effects due to small scales dynamics become relevant, the power spectra of the cosmic convergence (or shear) in the 
flat--sky approximation in UDM and $\Lambda$CDM are similar. When the Jeans' length $\lambda_J(a)$ increases, the Newtonian potential 
starts to decay earlier in time (at a fixed scale), or at greater scales (at a fixed epoch). This behaviour reflects on weak lensing by suppressing 
the convergence power spectra at high multipoles. They find that, for values of the sound speed between $c_\infty=10^{-3}$ and $c_\infty=10^{-2}$, 
UDM models are still comparable with $\Lambda$CDM, while for higher values of $c_\infty$ these models are ruled out because of the inhibition 
of structure formation. Moreover, they find that the dependence of the UDM weak lensing signal on the sound speed $c_\infty$ increases with 
decreasing redshift of the sources. They also show the errors for the fiducial $\Lambda$CDM signal for wide--field surveys like EUCLID or
Pan--STARRS, and they find that one isin principle able to distinguish $\Lambda$CDM from UDM models when $c_\infty\gtrsim10^{-2}$. 
Moreover, in Ref.~\cite{Camera:2010wm} the authors calculate the 3D shear matrix $C^{\gamma\gamma}(k_1,k_2;\ell)$ in the flat-sky 
approximation for a large number of values of $c_\infty$. They see that, whilst the agreement with the $\Lambda$CDM model is good 
for small values of $c_\infty$, when one increases the sound speed parameter, the lensing signal appears more suppressed at small scales, 
and the 3D shear matrix shows bumps related to the oscillations of the gravitational potential. Moreover, they show that the 
expected evidence clearly shows that the survey data would unquestionably favour UDM models over the standard $\Lambda$CDM model, 
if its sound speed parameter exceed $10^{-4}$.

\subsection{Prescription for UDM Models with a generic kinetic term}
\label{UDMg(X)-generic}

We now describe a general prescription to obtain a collection of models
that reproduce a background similar to $\Lambda$CDM and have a suitable sound speed. 
Some comments about the master equation (\ref{mastereq}) are first necessary.
The relation (\ref{mastereq}) enables to determine a connection
between the scalar factor $a$ and the kinetic term $X$ on the 
manifold $\mathcal{M}_\Lambda$ and therefore a mapping between 
the cosmic time and the manifold $\mathcal{M}_\Lambda$.

Now it is easy to see that the LHS of Eq.~(\ref{mastereq}), seen as a single function of $X$,
must have at least a vertical asymptote and a zero, and the function must be continuous between the two. 
In particular, when $X$ is near the vertical asymptote the universe approaches the cosmological 
constant regime, whereas when $X$ is close to the zero of the function,
the dark fluid behaves like dark matter. 
Therefore, if we define
\begin{equation}
f(\varphi(X))=\frac{\mathcal{F}(X)}{2X(\partial g(X)/\partial X)}
\end{equation}
where, for example,
\begin{equation}
\mathcal{F}(X)=\frac{1}{\mu}\frac{X_f-X}{X-X_i}\;,
\end{equation}
(where $\mu$ is an appropriate positive constant)
the value of $X_f$ and $X_i$ are the zero and the asymptote mentioned above,
namely, when $a \rightarrow 0$ we have $X \rightarrow X_i$ and
when $a \rightarrow \infty$ we have $X \rightarrow X_f$.
Moreover, if $X_f>X_i$ we have $dX/dN>0$, whereas if $X_f<X_i$ 
we have $dX/dN<0$.
In other words, according to Eq.(\ref{mastereq}), 
\begin{equation}
X(a)=X_f \frac{1+(X_i/X_f)\Lambda \mu \nu ~a^{-3}}
{1+\Lambda \mu \nu ~a^{-3}}\;.
\end{equation}
Let us emphasize that the values of $X_i$ and $X_f$ are very important
because they automatically set the range of values 
that the sound speed can assume at the various cosmic epochs.

Let us finally make another important comment. One can use
this reconstruction of the UDM model in the opposite way.
In fact, by imposing a cosmological background
identical to $\Lambda$CDM, the observed CMB power spectrum, 
and the observed evolution of cosmic structures, one can derive the 
evolution of the sound speed $c_s^2$ vs. cosmic time.
In this case, by assuming an appropriate kinetic term $g(X)$ through Eq.~(\ref{c_s^2_fg-v}), 
we can derive $X(a)$ and, consequently,
$\varphi(a)$ and $X(a(\varphi))=X(\varphi)$. Therefore, by using the
relations (\ref{mastereq}) and $V(\varphi)=f(\varphi)g(X)+\Lambda$,
one can determine the functional form of $f(\varphi)$ and $V(\varphi)$.

\section{Generalized UDM Models}
\label{GUDM}

In this Section we consider several possible generalizations of the technique 
introduced in Section \ref{UDM}, with the aim of  
studying models where the background does not necessarily mimic
the $\Lambda$CDM background (see \cite{Bertacca:2008uf, Bertacca:2010mt}).

Let us consider a scalar field Lagrangian $\mathcal{L}$ of the form 
\begin{equation}
 \label{LagrangianGen}
 \mathcal{L}(X,\varphi)= p(\varphi ,X)=f(\varphi)g(h(\varphi)X)-V(\varphi) \; .  
\end{equation}
Note that, introducing the three potentials $f(\varphi)$, $h(\varphi)$ and $V(\varphi)$,
we follow an approach similar to the one studied in Ref.\ \cite{Bertacca:2008uf} in order to decouple the equation of state parameter $w$ and the sound speed $c_{\rm s}$.  
In order to reconstruct these potentials we need three dynamical conditions: $a)$ a choice for $p(N)$, $b)$ the continuity equation or, equivalently, the equation of motion (\ref{eq_phi}), $c)$ a choice for $c_s^2(N)$ (see \cite{Bertacca:2010mt}).

Let us obtain the Lagrangian through two different simple approaches:\\
\begin{itemize} 
\item[1)]By choosing $p(N)$. Indeed we get
\begin{equation}
\frac{d\rho}{dN}+3\rho=-3p(N)\;,\quad\quad{\rm i.e.}
\quad\quad \rho(N)=e^{-3N}\left[-3\int^N\left(e^{3N'}p(N')dN'\right) + 
K\right]\;,
\end{equation}
where $K$ is an integration constant.
By imposing the condition $\mathcal{L}(X,\varphi)=p(N)$ along 
the classical trajectories, we obtain 
$\varphi=\mathcal{L}^{-1}(X(N),p(N))\big|_{\mathcal{M}_{p(N)}}$.
Thus, starting from a generic Lagrangian $\mathcal{L} = 
f(\varphi)g(h(\varphi)X)-V(\varphi)$ we get
\begin{equation}
\label{gen2}
2X(N)\left[\frac{\partial g(h(\varphi[X,N])X)}{\partial X}\right](N)f(\varphi(X,N)) = 
p(N)+e^{-3N}\left[-3\int^N\left(e^{3N'}p(N')dN'\right)+K\right]\;.
\end{equation}
For example, if $p=-\Lambda$, $K=\rho(a=1)-\Lambda$. The freedom provided by
the choice of $K$ is particularly relevant. In fact,
by setting $K=0$, we can remove the term $\rho \propto a^{-3}$. Alternatively, 
when $K\ne 0$, we always have a term that behaves like presseure-less matter.
We thus show that the single fluid of UDM models can mimic
not only a cosmological constant but also any quintessence fluid.

Thus, using  Eq.~(\ref{gen2}) and following the procedure described 
in Section \ref{UDM}, one gets the relations
$X\equiv\mathcal{G}_p(N)$, and consequently 
\begin{eqnarray}
\varphi & \equiv & \mathcal{Q}_p (N) = \varphi_0 \nonumber \\
 & \pm & 
\int^N\left\{\mathcal{G}_p (N')^{1/2}
\left[-3e^{-3N}\int^N\left(e^{3N'}p(N')dN'\right) + 
Ke^{-3N}\right]^{-1/2}dN'\right\}\;.
\end{eqnarray}
Therefore, with the functions $\mathcal{G}_p (N)$ and $\mathcal{Q}_p (N)$, 
one can write 
$f(X,N)=f(\mathcal{G}_p(N),N)=
f(\mathcal{G}_p(\mathcal{Q}_p^{-1}(\varphi)),\mathcal{Q}_p^{-1}(\varphi)) = 
f(\varphi)$.
Thus, by starting from a Lagrangian whose behavior is given  
by $p(N)$, the speed of sound is determined by the appropriate 
choice of $g(h(\varphi)X)$, where $h(X,N)=h(\mathcal{G}_p(N),N)=
h(\mathcal{G}_p(\mathcal{Q}_p^{-1}(\varphi)),\mathcal{Q}_p^{-1}(\varphi)) = h(\varphi)$. 

\item[2)]By choosing the equation of state $w(N)$. Indeed
\begin{equation}
\rho(N)=\rho_{0}e^{-3\int^N(w(N')+1)dN'}\;,
\end{equation}
where $\rho_0$ is a positive integration constant, and
\begin{equation}
p(N)=\rho_{0}w(N)e^{-3\int^N(w(N')+1)dN'}\;.
\end{equation}
Therefore, still by imposing the condition $\mathcal{L}(X,\varphi) = 
p[w(N),N]$ along 
the classical trajectories, i.e.
$\varphi=\mathcal{L}^{-1}[X(N),p(w(N),N)]\big|_{\mathcal{M}_{w(N)}}$, one gets
\begin{equation}
\label{gen3}
2X\frac{\partial g(h(\varphi[X,N])X)}{\partial X}f(X,N)=\rho_0[w(N)+1]
e^{-3\int^N(w(N')+1)dN'}\;.
\end{equation}
Therefore, on the classical trajectory we can impose, by using $w(N)$, 
a suitable function $p(N)$ and thus the function $\rho(N)$.
The master equation Eq.~(\ref{gen3}) generalizes Eq.~(\ref{mastereq}).
Also in this case, by  Eq.~(\ref{gen3}) and by following the 
argument described in Section \ref{UDM}, one can get the relations
$X\equiv\mathcal{G}_w(N)$, and consequently 
\begin{equation}
\varphi\equiv\mathcal{Q}_w(N)=\pm \int^N\left\{\mathcal{G}_w(N')^{1/2}
\left[\rho_{0}e^{-3\int^{N'}(w(N'')+1)dN''}\right]^{-1/2}dN'\right\} + 
\varphi_0\;.
\end{equation}
Thus, with the functions $\mathcal{G}_w(N)$ and $\mathcal{Q}_w(N)$, 
one can write 
$f(X,N)=f(\mathcal{G}_w(N),N)=
f(\mathcal{G}_w(\mathcal{Q}_w^{-1}(\varphi)),\mathcal{Q}_w^{-1}(\varphi))
= f(\varphi)$.
Then we can find a Lagrangian whose behavior is determined  
by $w(N)$ and whose speed of sound is determined by the appropriate 
choice of $g(h(\varphi)X)$, where $h(X,N)=h(\mathcal{G}_p(N),N)=
h(\mathcal{G}_p(\mathcal{Q}_p^{-1}(\varphi)),\mathcal{Q}_p^{-1}(\varphi)) = h(\varphi)$. 
\end{itemize}

Let us conclude that the $p(N)$ constraint on the 
equation of motion is actually a  weaker condition than the $w(N)$ 
constraint. The larger freedom that the $p(N)$ constraint provides
naturally yields an additive term in the energy density that decays
like $a^{-3}$, i.e. like a matter term 
in the homogeneous background. 
Let us emphasize that this important result is a natural consequence of the 
$p(N)$ constraint and is not imposed {\it a priori} (see \cite{Bertacca:2008uf, Bertacca:2010mt}). 

\section{Halos of Unified Dark Matter Scalar Field}\label{HUDM}

A complete analysis of UDM models should necessarily include 
the study of static solutions of Einstein's field
equations. This is complementary to the study of cosmological background
solutions and would allow to impose further constraints to the Lagrangian 
of UDM models.
The authors of Refs.~\cite{ArmendarizPicon:2005nz} and 
\cite{DiezTejedor:2006qh} have studied spherically symmetric and 
static configuration for k-\emph{essence} models. 
In particular, they studied models where the rotation velocity becomes flat 
(at least) at large radii of the halo.
In these models the scalar field pressure is not small 
compared to the mass-energy density, similarly to what is found in the study 
of general fluids in
Refs.~\cite{Bharadwaj:2003iw,Lake:2003tr,Faber:2005xc,Faber:2006sb}, 
and the Einstein's equations of motion do not reduce to the equations 
of Newtonian gravity. 
Further alternative models have been considered, even 
with a canonical kinetic term in the Lagrangian, that
describe dark matter halos in terms of bosonic scalar fields,
see e.g. Refs.~\cite{Lee:1995af,Wetterich:1994bg,Wetterich:2001cn,Wetterich:2001yw,Sahni:1999qe, Arbey:2001jj,Arbey:2003sj,Arbey:2006it,Matos:1998vk,Bilic:2005sn}.\\
In this Section we assume that our scalar field 
configurations only depend on the radial direction.  
Three main results are achieved. First, we are able to find a purely kinetic
Lagrangian which allows simultaneously to provide a flat rotation curve and
to realize a unified model of dark matter and dark energy 
on cosmological scales.
Second, an invariance property of the expression for the halo 
rotation curve is found. This allows to obtain purely kinetic Lagrangians that reproduce
the same rotation curves that are obtained starting from a given 
density profile 
within the standard Cold Dark Matter (CDM) paradigm. 
Finally, we consider a more general class 
of models with non-purely kinetic Lagrangians. In this case one can extend to 
the static and spherically symmetric spacetime metric the procedure used 
in Ref.~\cite{Bertacca:2007ux} to find UDM solutions in a cosmological setting.
Such a procedure requires that the Lagrangain is constant 
along the classical trajectories; we are thus able 
to provide the conditions to obtain reasonable rotation curves 
within a UDM model of the type discussed in Ref.~\cite{Bertacca:2007ux}. 

\subsection{Static solutions in Unified Dark Matter models}

Let us consider a scalar field which is static and spatially inhomogeneous, i.e.
such that $X<0$. 
In this situation the energy-momentum tensor is not described
by a perfect fluid and its stress energy-momentum tensor reads 
\begin{equation}
T^{\varphi}_{\mu \nu } = (p_\parallel+\rho) n_\mu n_\nu - \rho g_{\mu \nu }
\end{equation}
where 
\begin{equation}
\label{energy-density}
\rho=-p_{\perp}=-\mathcal{L}\;,
\end{equation}
$n_\mu=\nabla _{\mu }\varphi/\sqrt{-2X}$
and $p_{\parallel}=\mathcal{L} - 2X\partial \mathcal{L}/\partial X $.
In particular, $p_{\parallel}$ is the pressure in the direction parallel to 
$n_\mu$ whereas $p_{\perp}$ is the pressure in the direction orthogonal to 
$n_\mu$. It is simpler to work with a new definition of $X$. 
Indeed, defining $X=-\chi$ we have
\begin{eqnarray}
\label{def_n_with_chi}
n_\mu=\nabla _{\mu }\varphi/(2\chi)^{1/2}\\ 
\label{def-p_with_chi}
p_{\parallel}=2\chi\frac{\partial\rho}{\partial\chi} - \rho \;. 
\end{eqnarray}
Let us consider for simplicity the general static spherically 
symmetric spacetime metric i.e.
\begin{equation}
ds^2=-\exp{(2\alpha(r))}\;dt^2+\exp{(2\beta(r))}\;dr^2+r^2 d\Omega^2\, ,
\end{equation}
where $d\Omega^2=d\theta^2+\sin^2 \theta d\phi^2$ and $\alpha$
and $\beta$ are two functions that only depend upon $r$.\\
As the authors of Refs.~\cite{ArmendarizPicon:2005nz,DiezTejedor:2006qh} have 
shown, it is easy to see that the non-diagonal term $T^{rt}$ vanishes. 
Therefore $\varphi$ could be either 
strictly static or depend only on time. In this section we study the
solutions where $\varphi$ depends on the radius only.\\

In the following we will consider some cases where the 
baryonic content is not negligible in the halo. In this case  
we will assume that most of the baryons are concentrated 
within a radius $r_b$. If we define $M_*$ as the entire mass 
of the baryonic component then for $r>r_b$ we can simply assume that $M_*$ 
is concentrated in the center of the halo. 

Considering, therefore, the halo for $r>r_b$,
starting from the Einstein's equations and the covariant conservation of
the stress-energy (or from the equation of
motion of the scalar field, Eq.~(\ref{eq-motion})), 
we obtain 
\begin{eqnarray}
\label{eqE1}
\frac{1}{r^2}\left\{1-\left[r\exp{\left(-2\beta\right)}\right]'\right\}=\rho
&\Longleftrightarrow& \frac{dM}{dr}=4\pi\rho r^2 \;,\\
\label{eqE2}
\frac{1}{r^2}\left\{\exp{\left[-2(\alpha+\beta)\right]}
\left[r\exp{\left(2\alpha\right)}\right]'-1\right\}=p_{\parallel}
&\Longleftrightarrow& \alpha'=
\frac{\frac{M+M_*}{8\pi}+\frac{p_{\parallel}r^3}{2}}
{r^2\left[1-\frac{M+M_*}{4\pi r}\right]} \;, \nonumber\\
\\
\label{eqE3}
\frac{\exp{\left[-(\alpha+2\beta)\right]}}{r}
\left\{\left[r\exp{\alpha}\right]'\beta'-
\left[r\left(\exp{\alpha}\right)'\right]'\right\}=\rho \;,
\end{eqnarray}
\begin{equation}
\label{eq-conserv-st}
\frac{dp_\parallel}{dR}=-(p_\parallel+\rho)
\end{equation}
\noindent
(which are the $00$, $rr$ and $\theta\theta$ components of Einstein's 
equations and the $r$ component of the continuity equation respectively)
where 
\[\exp{(-2\beta(r))}=1-(M+M_*)/(4\pi r)\] and
\[R=\ln[r^2\exp(\alpha(r))]\] \;, 
where a prime indicates differentiation with 
respect to the radius $r$.\\ 

A first comment is in order here. If {\it i)} 
$\beta'=0$ and {\it ii)} $\left[r\left(\exp{\alpha}\right)'\right]'>0$,  
then we can immediately see that $\rho<0$. These conditions must therefore 
be avoided when trying to find a reasonable rotation curve. For example,
neglecting the baryonic mass, the special case of $\rho=A/r^2$ and 
$\exp(\alpha)\sim r^m$, where $A$ and $m$ are constants, fall into this case.
One thus recovers the \emph{no-go} theorem derived in 
Ref.~\cite{DiezTejedor:2006qh} under the assumption  
that the rotation curve $v_c\ll 1$ is constant for all $r$.\\
The value of the circular velocity $v_c$ is determined by the assumption 
that a massive test particle is also located at $\theta=\pi/2$. 
We define as massive test particle the object that
sends out a luminous signal to the observer who is considered to be 
stationary and far away from the halo.

Considering the motion of a massive test particle, say a star, in a such a halo, 
its trajectory is then described by a curve $x^{\mu}=(t,r,\theta,\phi)$ 
parameterized by some affine parameter; here we use its proper time $\tau$. 
Its four velocity is then simply $u^\mu\equiv dx^\mu/d\tau$. 
Due to spherical symmetry, we can assume without loss of generality that 
the star's ecliptic is located in the $\theta=\pi/2$ plane. 
Since the star is a massive particle, its norm is $u_\mu u^\mu=-1$, 
which becomes the constraint equation
\begin{equation}\label{eq:normalization}
	\exp{(2\alpha)}\dot{t}^2-\exp{(2\beta)}\dot{r}^2-r^2\dot{\phi}^2=1,
\end{equation}
where a dot denotes a derivative with respect to proper time $\tau$. Since the metric 
does not explicitly depend on $\theta$, the star's angular momentum $l$ is conserved,
\begin{equation}\label{eq:l}
	l=r^2\dot{\phi}.
\end{equation} 
Similarly, the metric does not explicitly depend on $t$, and there is a conserved energy $E$,
\begin{equation}\label{eq:E}
	E=\exp{(2\alpha)}\dot{t}.
\end{equation}
Substituting equations (\ref{eq:E}) and (\ref{eq:l}) into equation (\ref{eq:normalization}), 
one finds a first integral of motion for the star,
\begin{equation}
	\frac{1}{2}\dot{r}^2+\mathcal{V}(r)=0, \label{eq:dop1}
\end{equation}
where its effective potential is
\begin{equation}  
	\mathcal{V}(r)=\frac{1}{2}\exp{(-2\beta)}\left(1+\frac{l^2}{r^2}\right)-\frac{1}{2}E^2 \exp{[-2(\alpha+\beta)]}. 
		\label{eq:dop2}
\end{equation}
Note that the potential explicitly depends on the energy. Stationary orbits at radius $r$ exist if $\mathcal{V}$ 
and $d\mathcal{V}/dr$ vanish at that radius.  The former condition yields 
\begin{equation}
	1+\frac{l^2}{r^2}=E^2 \exp{[-2\alpha(r)]},
	\label{eq:dopcond1}
\end{equation}
whereas  the latter gives us
\begin{equation}
	-\beta'\left(1+\frac{l^2}{r^2}\right)
	-\frac{l^2}{r^3}+E^2\left(\alpha'+\beta'\right)\exp{[-2\alpha(r)]}=0
	\label{eq:dopcond2}.
\end{equation}
Substituting equation (\ref{eq:dopcond1}) into equation (\ref{eq:dopcond2}) and using the 
Eqs.~(\ref{eqE1}), (\ref{eqE2}) and (\ref{eq-motion}), we get the following equation 
\begin{equation}
	\frac{l^2/r^2}{1+l^2/r^2}=\frac{(M+M_*)(r)}{8\pi r}+ r^2\, \frac{p_\parallel(r)}{2},
	 \label{eq:doppler}
\end{equation}
which directly relates the angular momentum $l$ to the density profile of the halo. 

In this case, through the definition of the star's angular momentum $l$ and Eq.~(\ref{eq:doppler}), 
the value of $v_c\equiv l/r$ can be rewritten as 
\begin{equation}
\label{v_c}
v_c^2=\frac{p_\parallel r^2/2+(M+M_*)/(8\pi r)}
{1-\left[p_\parallel r^2/2+(M+M_*)/(8\pi r)\right]} \;, 
\end{equation}
but when we consider the weak-field limit condition 
$(M+M_*)/(8\pi r)\ll 1$ and since the rotation velocities of the halo of a 
spiral galaxy are typically non-relativistic, $v_c \ll 1$, 
Eq.~(\ref{v_c}) simplifies to \cite{ArmendarizPicon:2005nz}
\begin{equation}
\label{v_c-2}
v_c^2 \approx \frac{M+M_*}{8\pi r}+\frac{p_\parallel r^2}{2}\;.
\end{equation}

A second comment follows from the fact that the pressure is not small
compared to the mass-energy density. In other words we do not require
that general relativity reduces to Newtonian gravity (see also 
Refs.~\cite{Bharadwaj:2003iw,Lake:2003tr,Faber:2005xc,Faber:2006sb}). 
Notice also that in the region where 
$v_c \approx {\rm const.} \ll 1$ it is easy to see that in general 
$\exp(\alpha) \approx {\rm const.}$ since from Eqs.~(\ref{eqE2}) 
and~(\ref{v_c-2}) one obtains $r \alpha' \approx v_c^2$.
 
Finally, let us point out one of the main results (see also Ref.\  \cite{Bertacca:2008uf}). 
We can see that the relation (\ref{v_c-2}) 
is invariant under the following transformation
\begin{equation}
\label{transformation}
\rho\longrightarrow\widetilde{\rho}=\rho+\sigma(r) \quad\quad\quad
p_\parallel\longrightarrow\widetilde{p_\parallel}=p_\parallel+q(r)
\end{equation}
if  
\begin{equation}
\label{cond-q-sigma}
3q(r)+rq(r)'=-\sigma(r)\;,
\end{equation}
up to a proper choice of some integration constants.
Thanks to this transformation we can consider an ensemble of solutions that 
have the same rotation curve. Obviously, these solutions have to
satisfy Einstein's equations (\ref{eqE1}), (\ref{eqE2}) and (\ref{eqE3}),
and the covariant conservation of the stress-energy 
(\ref{eq-conserv-st}). 
Moreover, we will require the validity of the weak energy conditions, 
$\rho \geq 0$ and $p_\parallel+\rho \ge 0$, i.e. 
\begin{equation}
2\frac{\exp{\left(-2\beta\right)}}{r}(\alpha'+\beta')=
 2\chi\frac{\partial\rho}{\partial\chi}\ge 0\;.
\end{equation}

\subsection{Unified Dark Matter models 
with purely kinetic Lagrangians}

Let us consider a scalar field Lagrangian $\mathcal{L}$ 
with a non-canonical kinetic term that depends only on $X$ or $\chi$.   
Moreover, in this section we assume that 
$M_*=0$ (or $M\gg M_*$).

First of all we must impose that $\mathcal{L}$ is negative when $X<0$, so that
the energy density is positive. Therefore, we define a new positive function 
\begin{equation}
g_s(\chi)\equiv - \mathcal{L}(X)\;.
\end{equation}
As shown in Ref.~\cite{ArmendarizPicon:2005nz}, when
the equation of state $p_\parallel=p_\parallel(\rho)$ is known, one 
can write the purely kinetic Lagrangian that describes this dark fluid 
with the help of Eqs.~(\ref{energy-density}) and 
(\ref{def-p_with_chi}). 
Alternatively, using (\ref{eq-conserv-st}), one can connect 
$p_\parallel$ and $\rho$ in terms of $r$ through the variable $R$. 
Moreover, it is easy to see that 
starting from the field equation of motion (\ref{eq-motion}), 
there exists another relation that connects $\chi$ (i.e. $X$) with $r$.
This relation is  
\begin{equation}
\label{gs-chi-r}
\chi \left[\frac{dg_s(\chi)}{d\chi}\right]^2=
\frac{k}{\left[r^2\exp{\alpha(r)}\right]^2}
\end{equation}
with $k$ a positive constant. 
If we add an additive constant to $g_s(\chi)$, 
the solution (\ref{gs-chi-r}) remains unchanged.
One can see this also through Eq.~(\ref{eq-conserv-st}).
Indeed, using Eqs.~(\ref{energy-density}) and ~(\ref{def-p_with_chi})
one immediately finds that Eq.~(\ref{eq-conserv-st}) is invariant under
the transformation $\rho \rightarrow \rho + K$ \;
$p_\parallel \rightarrow p_\parallel - K$.
In this way we can add the cosmological constant
$K=\Lambda$ to the Lagrangian and we can describe the dark matter and the
cosmological constant-like dark energy as a single dark fluid i.e. 
as Unified Dark Matter (UDM). 

Let us notice that one can adopt two approaches to find 
reasonable rotation curves $v_c(r)$. A static solution  can be studied
in two possible ways: 
\begin{itemize} 
\item[i)]
The first approach consists simply in adopting directly
a Langrangian that provides a viable cosmological  
UDM model and exploring what 
are the conditions under which it can give a static solution 
with a rotation curve that is flat at large radii. This prescription has been 
already applied, for example, in Ref.~\cite{ArmendarizPicon:2005nz}.
\item[ii)]
A second approach consists in exploiting the invariance 
property of Eq.~(\ref{v_c-2}), 
with respect to the transformation (\ref{transformation}) (when the condition
(\ref{cond-q-sigma}) is satisfied). 
Usually in the literature one reduces the problem to the 
Newtonian gravity limit, because one makes use of a CDM density profile, 
i.e. one assumes that in 
Eq.~(\ref{v_c-2}), $p_\parallel \ll M/(4\pi r^3)$. We can therefore use 
Eqs.~(\ref{transformation}) and (\ref{cond-q-sigma})
to obtain energy density and pressure profiles $\rho(r)$ and 
$p_\parallel(r)$ that reproduce the same rotation curve in a model with 
non-negligible pressure. 
Next, we find an acceptable equation of state $p_\parallel=p_\parallel(\rho)$
such that we can reconstruct, through Eqs.~(\ref{energy-density}) and 
(\ref{def-p_with_chi}), the expression for the Lagrangian $\mathcal{L}$. 
Such a procedure establishes a mapping between UDM and CDM solutions 
that predict the same halo rotation curve $v_c(r)$.
As a starting point we could, of course, use very different CDM density 
profiles to this aim, such as the modified isothermal-law profile
\cite{King:1962wi}, the Burkert profile \cite{Burkert:1995yz}, the Moore
profile \cite{Moore:1999gc}, the Navarro-Frenk-White profile 
\cite{Navarro:1996gj,Navarro:2003ew} or the profile proposed 
by Salucci et al. (see for example \cite{Salucci:2007et, Salucci:2007tm, Shankar:2006xz, Salucci}). \\
As we have already mentioned, the possible solutions one finds in 
this way have to
satisfy the Einstein equations (\ref{eqE1}), (\ref{eqE2}) and (\ref{eqE3}),
the conservation of stress-energy (\ref{eq-conserv-st}) and the weak energy 
conditions. Moreover, the resulting UDM scalar field Lagrangian must 
be able to provide cosmological solutions that yield an acceptable 
description of the cosmological background 
(see, e.g., Ref.~\cite{Bertacca:2007ux}) and low effective speed of sound 
(see for example Refs.~\cite{Garriga:1999vw,Mukhanov:2005sc,Bertacca:2007cv}) 
so that cosmic structure formation successfully takes place and 
CMB anisotropies fit the observed pattern 
\cite{Sandvik:2002jz,Giannakis:2005kr,Carturan:2002si,Amendola:2003bz}. 
\end{itemize}

Below, using approach i), we provide a worked example of a UDM model
with purely kinetic Lagrangian which is able to describe a flat halo 
rotation curve and then, using approach ii), we give a general 
systematic procedure to obtain a possible Lagrangian of UDM model 
starting from a given CDM density profile.

\subsubsection{Approach i): The generalized Scherrer solution}

Let us consider the generalized Scherrer solution models obtained in 
Ref.~\cite{Bertacca:2007ux} (see also Sect. \ref{genScherrer}). These models are described by
the following Lagrangian
\begin{equation}
\mathcal{L}= -\Lambda + g_n \left(X-\hat{X}\right)^n 
\end{equation}
where $g_n>0$ is a suitable constant and $n>1$. 
The case $n=2$ corresponds to the unified model proposed by Scherrer
~\cite{Scherrer:2004au}.  
If we impose that today $[(X-\hat{X})/\hat{X}]^n \ll 1$, 
the background energy density can be written as 
\begin{equation}
\label{rho_lambda_dm}
\rho(a(t)) = \rho_\Lambda + \rho_{\rm DM} \;, 
\end{equation}
where $\rho_\Lambda$ behaves like a ``dark energy'' component 
($\rho_\Lambda = {\rm const.}$) and 
$\rho_{\rm DM}$ behaves like a ``dark matter'' component 
i.e. $\rho_{\rm DM}\propto a^{-3}$, with $a(t)$ the scale factor.\\
A static solution for the generalized Scherrer model can be obtained 
in two possible ways:
\begin{itemize} 
\item[1)] Starting from the analysis of Ref.~\cite{DiezTejedor:2006qh}, 
in the case of a barotropic Lagrangian for the homogeneous field.
The authors of Ref.~\cite{DiezTejedor:2006qh} indeed concluded that 
for $n \gg 1$ flat halo rotation curves can be obtained. 
In particular they studied spherically symmetric 
solutions with the following metric, 
\begin{equation}
ds^{2}=-\left(\frac{r}{r_{\star}}\right)^{b}dt^{2}+N(r)dr^{2}+
r^{2}d\Omega^{2}.\label{eq:flat}
\end{equation}
where $r_{\star}$ is a suitable length-scale and $b=2v_{c}^2$.
In the trivial case where $N(r)$ is constant they find $\mathcal{L}(X)\propto 
X^{2/b}$ with $b \ll 1$.
For $X \gg \hat{X} $ the Lagrangian 
$\mathcal{L}= -\Lambda + g_n (X-\hat{X})^n $ takes precisely this form.
\item[2)] In the analysis of Ref.~\cite{ArmendarizPicon:2005nz},  
solutions where $\varphi$ is only a function of the radius are considered. 
When the Lagrangian has the form $\mathcal{L}\propto X^n $, with $n \sim 10^6$ 
the halo rotation curve becomes flat at large radii. 
In this case $n$ must be an odd natural number, such that the energy density 
is positive. Our model is able to reproduce this situation when the matter 
density is large, i.e. when $|X| \gg \hat{X}$. 
\end{itemize}
Alternatively, if we wish to avoid large $n$ (c.f. case 2) above)
we can start from the following Lagrangian
\begin{equation}
\label{L-purelykessence}
\mathcal{L}= -\Lambda + \epsilon_X g_n  \left(|X|-\hat{X}\right)^n 
\end{equation}
where $\epsilon_X$ is some differentiable function of $X$ that is  
$1$ when $X\ge\hat{X}$ and $-1$ when $X\le-\hat{X}<0$. 
In this way when $X>\hat{X}>0$ we recover the Lagrangian of the generalized 
Scherrer solutions. When $X<0$ and $\chi=-X>\hat{X}$ we get
\begin{equation}
\mathcal{L}= -\Lambda - g_n \left(\chi-\hat{X}\right)^n 
\end{equation}
and, with the help of Eqs.~(\ref{energy-density}) and 
(\ref{def-p_with_chi}), we obtain 
\begin{equation}
\rho=-p_{\perp}=-\mathcal{L}\;, \quad \quad \quad
p_{\parallel}=(2n-1)g_n \left(\chi-\hat{X}\right)^n+2ng_n\hat{X} 
\left(\chi-\hat{X}\right)^{n-1}-
\Lambda\;.
\end{equation}
Now, requiring that $\chi$ be close to $\hat{X}$ (i.e.
$\left(\chi-\hat{X}\right)\ll \hat{X}$) and
$2ng_n\hat{X}\left(\chi-\hat{X}\right)^{n-1} \gg O(\Lambda)$,
and starting from the relation (\ref{gs-chi-r}) 
that connects $\chi$ with $r$, we get
\begin{equation}
\label{gs-chi-r-scherrer}
\left(\chi-\hat{X}\right)^{n-1}=\frac{k^{1/2}}{n g_n \hat{X}^{1/2}}\;
\frac{1}{r^2\exp{(\alpha(r))}}\;.
\end{equation}
Consistency with our approximations implies that we have to consider 
the following expressions for radial configurations with 
$r$ bigger than a minimum radius $r_{min}$. 
In this case  $p_{\parallel}$ and $\rho$ become
\begin{equation}
p_{\parallel}=\frac{A}{r^2\exp{(\alpha(r))}} \;, \quad \quad \quad
\rho=\frac{B}{[r^2\exp{(\alpha(r))}]^{n/(n-1)}}
\end{equation}
where $A=2(k\hat{X})^{1/2}$ and 
$B=g_n \left[k^{1/2}/(n g_n \hat{X}^{1/2})\right]^{n/(n-1)}$.\\

Using Eqs.~(\ref{eqE1})~and~(\ref{eqE2}), one can 
calculate the values of the metric terms $\exp{(\alpha)}$ and $\exp{(\beta)}$
 and, thus the value of $\rho$ and $p_{\parallel}$. Alternatively we know that 
when $v_c\approx {\rm const. \ll 1}$ at large radii, in a first approximation,
we can set $\exp(\alpha(r))\approx 
C = $ const.  Therefore for $n \neq 3$, we can write 
the function $M$ as 
\begin{equation}
M(r)\approx \frac{4\pi B}{C^{n/(n-1)}}
\left(\frac{n-1}{n-3}r^{\frac{n-3}{n-1}}+D\right)
\end{equation}
where we could also set $D=0$ for $n>3$. 
Instead, when $1<n<3$, the second term has to be larger than the first one.

In these cases $v_{c}^2$ becomes 
\begin{equation}
v_{c}^2(r) \approx \frac{A}{2 C}+
\frac{B}{2\;C^{n/(n-1)}}\left(\frac{n-1}{n-3}
\frac{1}{r^{2/(n-1)}}+\frac{D}{r}\right)\;.
\end{equation}
For $n=3$ we have 
\begin{equation}
M(r)\approx \frac{4\pi B}{C^{3/2}}
\ln\left(\frac{r}{\bar{r}}\right)+M(\bar{r})
\end{equation}
where $r > \bar{r}$ and
\begin{equation}
v_{c}^2\approx \frac{A}{2 C}+
\frac{B}{2\;C^{3/2}}\frac{1}{r}
\ln\left(\frac{r}{\bar{r}}\right)+\frac{M(\bar{r})}{8\pi r}\;.
\end{equation}
In other words we see that the circular velocity becomes  
approximately constant for sufficiently large $r$. 

However, let us stress that $\exp(\alpha(r))$ cannot be strictly  
constant, and that it should be chosen in such a way that  
the positivity of Eq.~(\ref{eqE3}) is ensured. 

This example can be generalized also to $M_*\neq 0$. Obviously,  
in such a case we have to assume that $r>r_b\ge r_{min}$. 
In this case $k$, $r_{min}$, $A$, $B$ (through 
$\exp (\beta(r))$) and $C$ depend on $M_*$.

The spherical top-hat solution for this model, which provides the link with 
the cosmological initial conditions, is described in the Sect. \ref{Static_genScherrer}.

\subsection{Approach ii): 
A general prescription to obtain UDM Lagrangians starting 
from a profile of an energy density distribution of CDM}

Defining the energy density distribution of CDM as
$\rho_{\rm CDM}(r)$ (with $p_{\rm CDM}=0$), the transformation 
(\ref{transformation})
 becomes 
\begin{equation}
\label{transformation-CDM}
\rho(r)=\rho_{\rm CDM}(r)+\sigma(r)\, , \quad\quad\quad
p_\parallel(r)=q(r) \;.
\end{equation}
Now, starting from a given CDM density profile, through 
Eqs.~(\ref{eqE1}), (\ref{eqE2}), 
(\ref{eq-conserv-st}) and   (\ref{cond-q-sigma})
we can determine~$\exp{(\alpha)}$~, ~$\exp{(\beta)}$, 
~$\rho$ ~and ~$p_\parallel$. In a second step we provide the 
conditions to ensure that the energy density is positive 
\footnote{Thanks to this condition, through Einstein's Eq.~(\ref{eqE3}), 
we can evade the \emph{no-go} theorem derived in 
Ref.~\cite{DiezTejedor:2006qh}.}. 
In this case, after some simple but lengthy calculations, one finds
\begin{eqnarray}
\label{Q}
\mathcal{Q}'(r)\left(r\frac{M_{\rm CDM}(r)}
{4\pi}-2r\mathcal{Q}(r)\right)
&-&2\mathcal{Q}^2(r)\nonumber \\ 
+\mathcal{Q}(r)\left(4r+3\frac{M_{\rm CDM}(r)}{4\pi}+
4r^3\rho_{\rm CDM}\right)&=&
\frac{r M_{\rm CDM}(r)}{4\pi}\left(4
+3r^2\rho_{\rm CDM}\right)\;,\nonumber \\ 
\\
\label{B}
\mathcal{B}(r)&=&\mathcal{Q}(r)-\frac{M_{\rm CDM}(r)}{4\pi}\;,\\
\label{A}
\mathcal{A}(r)&=&\frac{\mathcal{Q}(r)+\mathcal{B}(r)}{2\mathcal{B}(r)}\;,\\
\sigma(r)&=&\frac{1-\mathcal{Q}'(r)}{r^2}
\end{eqnarray}
\noindent
where $\mathcal{Q}(r) = r(r^2q+1)$~, ~$\mathcal{B}(r)=r\exp{(-2\beta)}$
and $\mathcal{A}(r)=(r\alpha'+1)$. Here we define
$M_{\rm CDM}(r)=4\pi \int_{0}^r \tilde{r}^2
\rho_{\rm CDM}(\tilde{r})~d\tilde{r}$.
At this point it is easy to see that Eq.~(\ref{Q}) does not admit a simple
analytical solution for a generic $\rho_{\rm CDM}$.
On the other hand we know that, through $\rho_{\rm CDM}$, all these functions
depend on the velocity rotation curve $v_c(r)$. Moreover $v_c^2(r)\ll 1$. 
Therefore, defining $\bar{v}_c$ as the value that $v_c$ assumes when the 
rotation curve is flat at large radii or the maximum value of $v_c$
with a particular profile of $\rho_{\rm CDM}$,
one can expand $\mathcal{Q}$, $\mathcal{A}$ and $\mathcal{B}$ 
as 
\begin{eqnarray}
\label{Q-pet}
\mathcal{Q}(r)&=&\mathcal{Q}_{(0)}(r)+\bar{v}_c^2 \mathcal{Q}_{(1)}(r)
+\frac{\left(\bar{v}_c^2\right)^2}{2!} \mathcal{Q}_{(2)}(r)+ \dots \;,
\nonumber\\
\label{A-pet}
\mathcal{A}(r)&=&\mathcal{A}_{(0)}(r)+\bar{v}_c^2 \mathcal{A}_{(1)}(r)
+\frac{\left(\bar{v}_c^2\right)^2}{2!} \mathcal{A}_{(2)}(r)+ \dots\;,
\nonumber\\
\label{B-pet}
\mathcal{B}(r)&=&\mathcal{B}_{(0)}(r)+\bar{v}_c^2 \mathcal{B}_{(1)}(r)
+\frac{\left(\bar{v}_c^2\right)^2}{2!} \mathcal{B}_{(2)}(r)+ \dots\;. 
\end{eqnarray}
\noindent

Following this procedure one can determine $\rho$ and $p_\parallel$ 
in a perturbation way, i.e.
\begin{equation}
\rho(r)=\rho_{(0)}(r)+\bar{v}_c^2 \rho_{(1)}(r) + 
\frac{\left(\bar{v}_c^2\right)^2}{2!} \rho_{(2)}(r)+ \dots\;,
\end{equation}
\begin{equation}
p_\parallel(r)=p_{\parallel \;(0)}(r)+\bar{v}_c^2 p_{\parallel \;(1)}(r)
+\frac{\left(\bar{v}_c^2\right)^2}{2!} p_{\parallel \;(2)}(r)+ \dots\;.
\end{equation}
\noindent
Now, looking at the various CDM density profiles
which have been proposed in the literature 
\cite{King:1962wi,Burkert:1995yz,Moore:1999gc,Navarro:1996gj,Navarro:2003ew,Salucci}, we see that 
one can always take $\rho_{\rm CDM}$ as
\begin{equation}
\rho_{\rm CDM}(r)=\bar{v}_c^2 \rho_{{\rm CDM}\;(1)}(r)\;, 
\end{equation}
then
\begin{equation}
 M_{\rm CDM}(r)=\bar{v}_c^2 M_{{\rm CDM}\;(1)}(r)=
 4\pi ~\bar{v}_c^2 \int_{0}^r \tilde{r}^2\rho_{{\rm CDM}\;(1)}
(\tilde{r})~d\tilde{r}\;.
\end{equation}
\noindent
For the zeroth-order terms one immediately obtains 
\begin{eqnarray}
\label{Q-0pet}
\mathcal{Q}_{(0)}&=&r\;,\nonumber\\
\label{A-0pet}
\mathcal{A}_{(0)}&=&1\;,\nonumber\\
\label{B-0pet}
\mathcal{B}_{(0)}&=&r\;.
\end{eqnarray}
\noindent
At the first order one gets
\begin{eqnarray}
\label{Q-1pet}
\mathcal{Q}_{(1)}&=&\frac{2}{r}\int_{0}^r 
\tilde{r}^3\rho_{{\rm CDM}\;(1)}
(\tilde{r})~d\tilde{r}\;,\nonumber\\
\label{A-1pet}
\mathcal{A}_{(1)}&=&\frac{1}{2r} 
\frac{M_{{\rm CDM}\;(1)}(r)}{4\pi}\;,\nonumber\\
\label{B-1pet}
\mathcal{B}_{(1)}&=&\frac{2}{r}\int_{0}^r 
\tilde{r}^3\rho_{{\rm CDM}\;(1)}
(\tilde{r})~d\tilde{r}-\frac{M_{{\rm CDM}\;(1)}(r)}{4\pi} \;.
\end{eqnarray}
\noindent
For completeness we write also the second order for $\mathcal{Q}$
\begin{eqnarray}
\label{Q-2pet}
\mathcal{Q}_{(2)}=\frac{1}{r} \int_{0}^r d\check{r}
\frac{M_{{\rm CDM}\;(1)}(\check{r})}{4\pi}
\left[\frac{2}{\check{r}} \mathcal{Q}_{(1)}({\check{r}}) 
- {\check{r}}^2 \rho_{{\rm CDM}\;(1)}
({\check{r}}) \right] \,.
\end{eqnarray}
\noindent
Let us stress that if one considers also terms 
$O\left(\bar{v}_c^4\right)$, Eq.~(\ref{v_c})
instead of Eq.~(\ref{v_c-2}) should be used. 
In such a case, $v_c$ slightly changes
with respect to the velocity rotation curve that one obtains using a 
CDM density profile.

For our purposes we can consider only the zeroth and the first-order 
terms.
At this point, one can finally calculate the value of ~$\rho$ ~and
~$p_\parallel$. One gets
\begin{equation}
\rho(r)=\rho_{\rm CDM}(r)+\frac{1-\mathcal{Q}'(r)}{r^2}=
\bar{v}_c^2 \left(\frac{2}{r^4}\int_{0}^r 
\tilde{r}^3\rho_{{\rm CDM}\;(1)}
(\tilde{r})~d\tilde{r}-\rho_{\rm CDM\;(1)}(r)\right)\;,
\end{equation}
\begin{equation}
p_\parallel(r)=\frac{\mathcal{Q}(r)-r}{r^3}=\bar{v}_c^2
\frac{2}{r^4}\int_{0}^r \tilde{r}^3\rho_{{\rm CDM}\;(1)}
(\tilde{r})~d\tilde{r}\;.
\end{equation} 
\noindent
As far as the values of the metric terms $\exp{(\alpha)}$ and $\exp{(\beta)}$
are concerned, we obtain the following expressions
\begin{equation}
\exp{(2\alpha)}=\exp{(2\alpha(\hat{r}))}\exp{\left[\bar{v}_c^2\int_{\hat{r}}^r
\frac{1}{\tilde{r}^2}\frac{M_{{\rm CDM}\;(1)}
(\tilde{r})}{4\pi}
~d\tilde{r} \right]}
\end{equation} 
\begin{equation}
\exp{(-2\beta)}=1+\frac{\bar{v}_c^2}{r^2}\left(2\int_{0}^r 
\tilde{r}^3\rho_{{\rm CDM}\;(1)}
(\tilde{r})~d\tilde{r}-r \frac{M_{{\rm CDM}\;(1)}(r)}
{4\pi}\right)\;.
\end{equation}
\noindent
Now, it is immediate to see that if we want a positive energy density 
we have to impose $2~\int_{0}^r \tilde{r}^3\rho_{{\rm CDM}\;(1)}
(\tilde{r})~d\tilde{r} \ge r^4~\rho_{{\rm CDM}\;(1)}(r)$. From 
Eq.~(\ref{eqE1}) we know that $M(r)=4\pi \int_{\hat{r}_{(0)}}^r 
\tilde{r}^2\rho(\tilde{r})+ M(\hat{r}_{(0)})$ and
$M_{\rm CDM}(r)=4\pi \int_{\bar{r}}^r 
\tilde{r}^2\rho_{{\rm CDM}}(\tilde{r}) ~d\tilde{r}+M_{{\rm CDM}}(\bar{r})$. 
Therefore we need to know what is the relation 
between $\bar{r}$ and $\hat{r}_{(0)}$. 
This condition is easily obtained if we make use of 
Eq.~(\ref{v_c-2}). Indeed, we get
\begin{equation}\label{initial-cond}
\frac{M_{(1)}(\hat{r}_{(0)})-M_{{\rm CDM}\;(1)}(\bar{r})}{4 \pi}+
\frac{2}{\hat{r}_{(0)}}\int_{0}^{\hat{r}_{(0)}} 
\tilde{r}^3\rho_{{\rm CDM}\;(1)}(\tilde{r})~d\tilde{r}=
\int_{\bar{r}}^{\hat{r}_{(0)}}\tilde{r}^2
\rho_{{\rm CDM}\;(1)}(\tilde{r})~d\tilde{r}\; , 
\end{equation}
which finally guarantees the invariance of the rotation velocity with respect
to the transformation in Eqs.~(\ref{transformation}) and (\ref{cond-q-sigma}). 

Let us, to a first approximation, parametrize the various CDM 
density profiles, at very 
large radii (i.e. when we can completely neglect the baryonic component) as  
\begin{equation}
\rho_{\rm CDM}=\frac{\kappa~\bar{v}_c^2}{r^n}
\end{equation}
where $\kappa$ is a proper positive constant which depends on the particular 
profile that is chosen
~\cite{King:1962wi,Burkert:1995yz,Moore:1999gc,Navarro:1996gj,Navarro:2003ew,Salucci}. For example for many of the density 
profiles the slope is $n=3$ for large radii 
\cite{Burkert:1995yz,Moore:1999gc,Navarro:1996gj,Navarro:2003ew,Salucci}.    

In this case a positive energy density $\rho>0$ requires $n\ge 2$.
At this point let us focus on the case where $2 \le n<4$, since this gives 
rise to the typical slope of most of the 
density profiles studied in the literature.
Therefore we obtain  for $\rho(r)$ and $p_\parallel(r)$:
\begin{equation}
\label{rho-p_2<n<4}
\rho(r)=\bar{v}_c^2~\kappa\,\frac{n-2}{4-n}\frac{1}{r^n}\;,\quad\quad\quad
p_\parallel(r)=\bar{v}_c^2~\kappa\,\frac{2}{4-n}\frac{1}{r^n}\;.
\end{equation}
In particular,
\begin{itemize}
\item[1)]for $n=2$, we get
\begin{equation}
\label{rho-p_n=2}
\rho(r)=0,\quad\quad\quad
p_\parallel(r)=\rho_{\rm CDM}=\bar{v}_c^2~\kappa\,\frac{1}{r^2}\;,
\end{equation}
and for the relation between $\hat{r}_{(0)}$ and $\bar{r}$ one can choose, for example, $\hat{r}_{(0)}=\bar{r}=0$. 
In other words, for large radii we have that $\rho(r) \ll p_\parallel(r)$.
\item[2)]Also for $2<n<3$ one can choose $\hat{r}_{(0)}=\bar{r}=0$.
\item[3)]For $n=3$ 
\begin{equation}
\label{rho-p_n=2}
\rho(r)=\rho_{\rm CDM},\quad\quad\quad
p_\parallel(r)=\bar{v}_c^2~\kappa\,\frac{2}{r^3}\;,
\end{equation}
and, through Eq.~(\ref{initial-cond}), we have to impose that
\begin{equation}
\frac{M_{(1)}(\hat{r}_{(0)})-M_{{\rm CDM}\;(1)}(\bar{r})}{4 \pi}=
\ln{\left(\frac{\hat{r}_{(0)}}{\bar{r}}\right)}-2\;.
\end{equation}
Notice that the energy density profile is the same as the CDM one only for large radii so that  
$M_{(1)}(r)$ differs  from $M_{{\rm CDM}\;(1)}(r)$. 
\item[4)]In addition, for $3<n<4$, also through Eq.~(\ref{initial-cond}), 
we have to impose that
\begin{equation}
\frac{M_{(1)}(\hat{r}_{(0)})-M_{{\rm CDM}\;(1)}(\bar{r})}{4 \pi}=
\frac{\bar{r}^{3-n}}{n-3}-\frac{(n-2)}{(4-n)(n-3)}\hat{r}_{(0)}^{3-n}.
\end{equation}
\end{itemize}
Now let us focus where $2<n<4$. 
Starting from Eq.~(\ref{rho-p_2<n<4}) to express 
$p_\parallel=p_\parallel(\rho)$ we solve Eq.~(\ref{def-p_with_chi})
to recover the Lagrangian for the scalar field 
\begin{equation}
\label{rho-chi}
\rho(\chi)= - \mathcal{L}= k \chi^{\frac{n}{2(n-2)}}\;,\quad\quad\quad
p(\chi)=\frac{2k}{(n-2)}\chi^{\frac{n}{2(n-2)}}
\end{equation}
where $k$ is a positive integration constant.
We can see that, for this range of $n$, the exponent is larger 
than $1$; thus there are no problems with a possible instability
of the Lagrangian (see 
Refs.~\cite{ArmendarizPicon:2005nz,Babichev:2007dw,Rendall:2005fv}). 
Therefore, through the 
transformation $\rho \rightarrow \rho + \Lambda$ \;
$p_\parallel \rightarrow p_\parallel - \Lambda$, this
Lagrangian can be extended to describe a unified model of dark matter 
and dark energy.
Indeed, starting from the Lagrangian of the type (\ref{L-purelykessence}),
when $|X| \gg \hat{X}$ and if $k=g_n$, $\mathcal{L}$ takes precisely 
the form (\ref{rho-chi}).

Finally, we want to stress that this prescription does not apply only 
to the case of an adiabatic fluid, such as the one provided by 
scalar field with a purely kinetic Lagrangian, but it 
can be also used for more general Lagrangians 
$\mathcal{L}(\varphi,X)$.  

\section{Conclusions}\label{conclusion} 

In this work we explored the possibility that the 
dynamics of a single scalar field can account for a unified description 
of the Dark Matter and Dark Energy sectors, leading to a Unified Dark Matter (UDM) model.
In comparison with the standard DM + DE models (e.g.\ even the
simplest model, with DM  and a cosmological constant),  in
UDM models there are two simple but important aspects to consider: first, the fluid which 
triggers the accelerated expansion at late times is also the one which has to cluster
in order to produce the structures we see today. 
Second, from the last scattering to the present epoch, the energy density of the Universe 
is dominated by a single dark fluid, and therefore the gravitational potential 
evolution is determined by the background and perturbation evolution of just such a fluid. 
As a result the general trend is that the possible appearance of a sound speed 
significantly different from zero at late times corresponds to the appearance of a Jeans length (or a sound horizon) 
under which the dark fluid does not cluster any more, causing a strong evolution in time of the gravitational potential 
(which starts to oscillate and decay).
Specifically in this paper we have explored  UDM models defined by  Lagrangian of 
k-\emph{essence} models. This allows to find suitable solutions around 
which the scalar field describes a mixture of Dark Matter and Dark Energy.
Finally we also investigated the static and spherically symmetric 
solutions of Einstein's equations for a scalar field 
with non-canonical kinetic term.

\acknowledgments{D.B. would like to acknowledge the ICG Portsmouth for the hospitality during the development of this project and ``Fondazione Ing. Aldo Gini" for 
financial support. D.B research has been  partly supported by ASI  contract I/016/07/0 ``COFIS".  
The authors are grateful to D. Bacon, M. Bruni, S. Camera, A. Diaferio, T. Giannantonio, A. F. Heavens, T. D. Kitching, O. Piattella, 
D. Pietrobon, M. Pietroni, A. Raccanelli for fruitful collaborations about UDM models. The authors also thank  L.\ Amendola, B.\ Bassett, R.\ Crittenden, 
R.\ Maartens, S.\ Mollerach, M.\ Sasaki, S.\ Tsujikawa, M.\ Viel and D.\ Wands  for discussions and suggestions.}
\appendix

\section{Spherical collapse for generalized Scherrer solution models} \label{Static_genScherrer}

Let we assume a flat, homogeneous Friedmann-Robertson-Walker 
background metric.
In such a case, the background evolution of the Universe is characterized
completely by the following equations 
\begin{equation}
\label{eq_u1-back}
H^2 =\frac{1}{3} \rho\, ,
\end{equation}
\begin{equation}
\label{eq_u2-back}
\dot{H} = - \frac{1}{2}(p + \rho)\, ,
\end{equation}
where the dot denotes differentiation w.r.t. the cosmic time $t$.\\ 
Now let us consider a top-hat spherical over-density with the purely kinetic
model with the Lagrangian $\mathcal{L}= -\Lambda + g_n (X-\hat{X})^n $ and
with $g_n>0$. For this particular case
within the over-dense region we have a single dark fluid undergoing 
spherical collapse, which is described by the following equation 
\begin{equation}\label{qr2_lammda=0}
\frac{\ddot R}{R} = -\frac{1}{6} \left(\rho_{R} + 
3 p_{R} \right)
\end{equation}
where $R$, $\rho_{R}$ and $p_{R}$ are respectively the scale-factor, pressure 
and energy density of the over-dense region; 
$\rho_{R}$ and $p_{R}$ are defined by the following expressions 
\begin{equation}
\label{rho_overdensity}
\rho_{R} = \Lambda + 2 n g_{n} \hat{X} (X_R - \hat{X})^{n-1} + 
(2n-1) g_{n} (X_R-\hat{X})^{n}
\end{equation}
\begin{equation}
\label{p_overdensity}
p_R= g_R = -\Lambda+g_{c}(X_R-\hat{X})^{n}
\end{equation}
with $X_R=X(R)$ a function of time. 

The equation of motion is
\begin{equation}
\label{collapse_lambda=0}
 \left(\frac{\partial g_R}{\partial X_R}+
2X\frac{\partial^2 g_R}{\partial X_R^2}\right)\frac{dX_R} {dN_R}
+ 3 \left(2 X_R \frac{\partial g_R}{\partial X_R}\right) = 0 \;. 
\end{equation}
where $dN_R=dR/R$. The solution of Eq.~(\ref{collapse_lambda=0})
(for $\partial g_R/\partial X_R , X_R \neq 0$ ) is
\begin{equation}
\label{sol_coll}
X_R\left(\frac{\partial g_R}{\partial X_R}\right)^{2} =
  k_R R^{-6} \end{equation}
where we can choose $k_R=R_{ta}^{6} \left[ X_R 
\left(\frac{\partial g_R}{\partial X_R} \right)^{2} \right]_{ta}$, with 
$R_{ta}$ the value of $R$ at turnaround.
Replacing  Eq.~(\ref{p_overdensity}) in Eq.~(\ref{sol_coll}) we find
\begin{equation}
\label{sol_coll2}
X_R \; \left[n g_n (X_R-\hat{X})^{n-1}\right]^2=  k_R R^{-6}
\end{equation}
Using now the explicit expressions for $\rho_R$ and $p_R$ we arrive at the 
following set of equations 
\begin{eqnarray}
\label{collapse_gen-eq}
\frac{\ddot R}{R}  = -\frac{1}{3} \left[-\Lambda + 
n g_{n} \hat{X} (X_R-\hat{X})^{n-1} + (n+1) g_{n} 
(X_R-\hat{X})^{n}\right] \\
(X_R-\hat{X})^{2n-1} + \hat{X} (X_R-\hat{X})^{2(n-1)} = 
\frac{k_R} {n^2 g_{n}^{2}} R^{-6}.
\end{eqnarray}

For $(X_R-\hat{X})/\hat{X} \ll 1$ Eq.~(\ref{collapse_gen-eq}) 
becomes
\begin{equation}
\label{collapse_eq_Xc_near_Xco}
\frac{\ddot R}{R} = - \frac{1}{3} \left\{-\Lambda +  n g_{n} 
|X_{R_{ta}}-\hat{X}|^{n-1} (X_{R_{ta}}\hat{X})^{\frac{1}{2}} 
\left(\frac{R} {R_{ta}}
\right)^{-3} \right\} 
\end{equation}

We can now write all the equations that describe 
the spherical collapse
\begin{eqnarray}
\label{eq_final_a}
\left(\frac{\dot a}{a} \right)^{2}  & = 
& \frac{1}{3} \left(\rho_\Lambda +  \rho_{\rm DM}\right) \\
   \rho_\Lambda & = & \Lambda \\
\label{eq_final_rho_k_DM}
       \rho_{\rm DM} & = 
& 2 n g_n |X_{ta}-\hat{X}|^{n-1} (X_{ta}\hat{X})^{\frac{1}{2}} 
\left(\frac{a} {a_{ta}}\right)^{-3} \\
\label{eq_final_R}
\frac{\ddot R}{R} & = & - \frac{1}{6} \left(\rho_{R_{DM}} 
- 2\rho_{R_{\Lambda}}\right) \\
\rho_{R_{\rm DM}} & = & 2 n g_{n} |X_{ta}-\hat{X}|^{n-1} 
(X_{R_{ta}}\hat{X})^{\frac{1}{2}} \left(\frac{R} {R_{ta}}\right)^{-3}
\end{eqnarray}
where $a_{ta}=a(t_{ta})$. 

Following now the same procedure of Ref.~\cite{Wang:1998gt} 
we can define $x$ and $y$ 
\begin{eqnarray}
\label{}
 x & \equiv & \frac{a} {a_{ta}} \\
 y & \equiv & \frac{R} {R_{ta}} \; .
\end{eqnarray}
In this way we can redefine $\rho_{\rm DM}$ and $\rho_{R_{\rm DM}}$ such that
\begin{eqnarray}
\label{eq_final_xy}
\rho_{\rm DM} & = & \frac{3 H_{ta}^{2}\Omega_{\rm DM}(x=1)} 
{x^{3}} \\
 \rho_{R_{\rm DM}} & = & \zeta \frac{3 H_{ta}^{2}\Omega_{\rm DM}(x=1)} 
{y^{3}} 
\end{eqnarray}
where $\Omega_{\rm DM}$ is the (k-\emph{essence}) dark matter density parameter, 
and $\zeta = 
(\rho/\rho_{\rm DM})|_{x=1}$. Then Eqs.~(\ref{eq_final_a}) 
and (\ref{eq_final_R}) become
\begin{eqnarray}
\label{}
        \frac{d x}{d \tau}  & = 
& \left(x \Omega_{_{DM}}(x) \right)^{- \frac{1}{2}} \;, \\ 
\frac{d^{2} y}{d \tau^{2}}  & = & -\frac{1}{2y^{2}}\left[\zeta - 2y^{3} 
K_{\Lambda} \right] \;, \\
             \Omega_{_{DM}}(x)  & = & \left(1- 
\frac{1-\Omega_{_{DM}}(x=1)}{\Omega_{_{DM}}(x=1)} 
x^{3}\right)^{-1} \;, 
\end{eqnarray}
where $ d \tau = H_{ta} \sqrt{\Omega_{_{DM}}(x=1)}$ and 
$ K_{\Lambda} = \rho_{_\Lambda}/[3H_{ta}^2 
\Omega_{_{DM}}(x=1)]$. 

Defining $U$ as the potential energy of the over-density 
and using energy conservation between virialization and turnaround, 
\begin{equation} 
\left[ U+\frac{R}{2}\frac{\partial U}{\partial R}  
\right]_{vir}  =  U_{ta} \label{ec} \;, 
\end{equation}
we obtain 
\begin{equation}
\label{}
(1+q)y-2qy^{3} = \frac{1}{2}
\end{equation}
where
\begin{equation}
\label{}
q = \left(\frac{\rho_{\Lambda}}{\rho}\right)_{y=1} = 
\frac{K_{\Lambda}}{\zeta} \;, 
\end{equation}
in full agreement with Ref.~\cite{Lahav:1991wc}. 

\bibliographystyle{JHEP}
\bibliography{BFastUDM-scalar_field-2}
\end{document}